\documentclass[11pt]{article}
\usepackage[margin=0.5in]{geometry}
\usepackage{blindtext} 

\usepackage[]{} 
\usepackage[T1]{fontenc} 
\usepackage{microtype} 
\usepackage[noadjust]{cite}
 
\usepackage[english]{babel} 

\usepackage[hang, small,labelfont=bf,up,textfont=it,up]{caption} 
\usepackage{booktabs} 

\usepackage{enumitem} 
\setlist[itemize]{noitemsep} 

\usepackage{abstract} 

\usepackage{titlesec} 
\renewcommand\thesection{\Roman{section}} 
\renewcommand\thesubsection{\roman{subsection}} 
\titleformat{\section}[block]{\large\bfseries\centering}{\thesection.}{.5em}{} 
\titleformat{\subsection}[block]{\large}{\thesubsection.}{1em}{} 

\usepackage{fancyhdr} 
\usepackage{titling} 
\usepackage{hyperref} 
\usepackage{amsmath,amssymb}
\usepackage{cuted}
\usepackage{flushend}
\usepackage{graphicx}
\usepackage{subcaption}
\usepackage{float}

\pretitle{\begin{center}\LARGE\bfseries} 
\posttitle{\end{center}} 
\title{Thermodynamic geometry of pure Lovelock black holes}
\author{
{Mohammadreza Ebrahimi Khuzani},\thanks{\hyperref{r.ebrahimi@ph.iut.ac.ir}{}{}{r.ebrahimi@ph.iut.ac.ir}}\ \
\textnormal{Behrouz Mirza},\thanks{\hyperref{b.mirza@iut.ac.ir}{}{}{b.mirza@iut.ac.ir}} \ \ \textnormal{Mahnaz Tavakoli Kachi},\thanks{\hyperref{m.tavakoli1386@ph.iut.ac.ir}{}{}{m.tavakoli1386@ph.iut.ac.ir}}\\
\normalsize\textit{Department of Physics, Isfahan University of Technology, Isfahan 84156-83111, Iran}\\[1ex]
}
\date{} 
\renewcommand{\maketitlehookd}{%
\begin{abstract}
\begin{changemargin}{2cm}{2cm} 
In this paper, we study thermodynamic geometry for pure Lovelock black holes. The thermodynamics
scalar curvature contains information about the interaction of microstates
that might be repulsive or attractive. We obtain critical exponents and critical amplitudes
of scalar and extrinsic curvatures for small and large black holes for various dimensions.
We demonstrate that this model’s thermodynamic Ricci scalar scaling behavior
has a universal behavior for different dimensions. Moreover, we determine the order
of phase transition by using Ehrenfest’s equations and the Prigogine–Defay ratio. We
also consider the extended phase space and investigate the critical behavior of the pure
Lovelock black holes for various dimensions.
\end{changemargin}
\end{abstract}}
\begin{document}
\maketitle
\section{INTRODUCTION}
Recently, gravity theories, which include curvature terms with the higher order of
derivatives, have attracted physicists’ attention. Among them, Lovelock theory was
constructed based on dimensional extended Euler densities, and its field equations
do not contain more than second-order derivative terms. This theory is free of ghost
fields \cite{lovelock}. At first order in three and four dimensions, it reduces to Einstein gravity.
The second-order curvature terms, so-called Gauss–Bonnet theory, only appears for
$d>4$. Generally, the Lovelock theory is a useful formulation for the corrections in
higher orders of curvature at short distances. It is a valuable theory about higherorder
curvature effects and AdS/CFT \cite{Maldacena,Witten,Gubser}.

The black holes entropy is one of the most critical issues of gravitational theories\cite{sw1, bekens1,page}.
 Although the macroscopic quantities of black holes can be reached by the action, the microstructure of black holes is unsettled \cite{bardeen,bekens2}. In 1975, Weinhold introduced
a geometric formulation for thermodynamic systems. He used Hessian matrix
for the internal energy of a thermodynamic system (or equivalency black hole’s
mass) that is a function of entropy and some other thermodynamic variables\cite{weinhold}.
After that, Ruppeiner utilized fluctuation theory in 1979 and represented a different
thermodynamic metric, and considered entropy as a function of some extensive
variables\cite{Rupp,Rupp2}. Over the years, thermodynamic geometry has been used for different
thermodynamic systems\cite{gibbons,geoex,caigeo,zamani,hernando,medved,anyon,prof4,fractional,thermomet,prof3,prof5,prof1}. Anyon particles and also thermodynamic systems
with fractional statistics were studied using thermodynamic geometry \cite{anyon}. The exact
correspondence between phase transition points and thermodynamic scalar curvature
divergence was proved in Refs. \cite{correspon, hessian}. Also, thermodynamic geometry has
been used to study black holes and their critical phase structures \cite{zsh1,zsh2,zsh3,zsh4,zsh5,zsh6,zsh7,zsh8,zsh9,zsh10, 4d}.

Thermodynamic extrinsic and intrinsic curvatures were obtained for different
kinds of black holes in Ref.\cite{extrin}. Besides phase transition, thermodynamic scalar
curvature, $R$, provides important information about the microstructure of black
holes as a thermodynamic system. In Ref. \cite{Rupp}, it was proposed that $ R\sim\kappa\xi^{d}$ where
$\xi$ is the correlation length, $\kappa$ is a constant and $d$ is dimensions of the system.
Correlation length diverges$(\xi \sim t^\nu)$ when temperature of thermodynamic system
tends to critical value $(t\to0)$, where $\nu$ is a critical exponent and $t=1-T/T_c$, where $T$ and $T_c$ are temperature and critical temperature, respectively.

The pure Lovelock gravity has been explored during the last years \cite{pure1}. The stability
of static pure Lovelock black holes has been investigated in Ref. \cite{dadhich1} and it
was shown in even dimensions the pure Lovelock black holes are unstable. However,
Lovelock black holes with a cosmological constant $\lambda$ are stable. It was proved
that in odd dimensions the Lovelock–Riemann tensor vanishes for any vacuum
solution of the pure Lovelock gravity\cite{dadhich2}. By using Hamiltonian formalism, the gravitational
equations for pure Lovelock gravity were investigated in Refs. \cite{dadhich3,dadhich4}.
The dynamical structure of pure Lovelock gravity for $d > 4$ was investigated in
Ref. \cite{dadhich5} by using Hamiltonian formalism. Furthermore, the critical behavior of pure
Lovelock black holes was explored\cite{pureeos}. It was shown that the critical exponents and
the critical behavior of pure Lovelock black holes are the same as the van der Waals
fluid (VdW). The critical exponents of the charged AdS black holes were obtained
by the normalized thermodynamic curvatures, the findings is same as those of the
previous study of the van der Waals fluid\cite{mans_crit}.

We study the thermodynamic geometry of pure Lovelock black holes. We calculate
thermodynamic Ricci scalars and extrinsic curvatures for pure Lovelock black
holes in $d=6, \ 8, \ 10$ dimensions. We investigate the critical behavior and obtain
the critical exponents of pure Lovelock black holes by using a numerical method.
We find that the phase transition behavior of black holes is the same as the van
der Waals thermodynamic systems. Then we use Ehrenfest’s equations to identify
the order of phase transitions in pure Lovelock black holes. Also, we derive
thermodynamic geometry in the extended phase-space. We also investigated Gibbs
free energy of pure Lovelock black holes.

This paper is organized as follows. In Sec. \ref{sec:pure}, we briefly review d-dimensional
black hole solutions and thermodynamic properties in pure Lovelock gravity. Section \ref{sec:pureg} is about different aspects of thermodynamic geometry of pure Lovelock black
holes. We first derive extrinsic and scalar thermodynamic curvatures then explore
the critical behavior of pure Lovelock black holes. In Sec. \ref{sec:ehrenfest} , we analyze phase transition
points by Ehrenfest approach and check out whether pure Lovelock black
holes satisfy two Ehrenfest’s equation. Furthermore, in Sec. \ref{sec:exten}, we consider the
extended phase-space and investigate thermodynamic geometry and critical behavior
in addition to Gibbs free energy of pure Lovelock black holes. Finally, a summary
of the results is given in Sec.  \ref{sec:conc}. Details on Poisson brackets method can be found
in Appendix A.

\section{CHARGED BLACK HOLES IN PURE LOVELOCK GRAVITY THEORY}\label{sec:pure}
The charged Lanczos-Lovelock gravity action in d-dimensions is written in the following form \cite{lovelock}
\begin{equation}
\begin{aligned}
&S=\frac{1}{16\pi G}\int{\sqrt{-g}\ d^dx\left(\sum_{n=0}^{p}{\alpha_n\mathcal{L}_n\left(R,\ R_{\mu\nu\ },R_{\mu\nu\rho\sigma\ },F_{\mu\nu}\ \right)}\right)}\\
& \ \ =\frac{1}{16\pi G}\int\sqrt{-g}\ d^dx\left( \alpha_0\mathcal{L}_0+\sum_{n=0}^{p}\alpha_n\mathcal{L}_n-F_{\mu\nu}F^{\mu\nu}\right), \label{eq:action}
\end{aligned}
\end{equation}
where $F_{\mu\nu}=\partial_\mu A_\nu-\partial_\nu A_\mu$ is the tensor of the electromagnetic field and $\alpha_{n}$ are coupling constants. The maximum value of $p$  is equal to $d/2$, which indicates the  order of gravity. This gravity theory is constructed by Euler density Lagrangians, $\mathcal{L}_n$, which are functions of the curvature scalar and tensors.  Considering just one non-vanishing coupling constant we obtain pure Lovelock gravity theory. We consider $ \mathcal{L}_0=-2\Lambda _0 $ for cosmological constant, $\mathcal{L}_1$ corresponds to the Einstein-Hilbert action and $\mathcal{L}_2$ denotes Einstein-Gauss-Bonnet (EGB) gravity theory. 
$\mathcal{L}_n$ are functions of the curvature scalar and tensors which are defined as
\begin{flalign}
	&\mathcal{L}_1=R,\\
	&\mathcal{L}_2=R^2-4R^{\mu\nu}R_{\mu\nu}+R^{\mu\nu\rho\sigma}R_{\mu\nu\rho\sigma},
\end{flalign}
and for $\mathcal{L}_n$
\begin{equation}
	\mathcal{L}_{n}=2^{-n}\delta_{c_{1}d_{1}\ .\ .\ .\ c_{n}d_{n}}^{a_{1}b_{1}\ .\ .\ .a_{n}b_{n}} R_{\ \ \ \ a_{1}b_{1}}^{c_{1}d_{1}}.\ .\ .\ .R_{\ \ \ \ a_{n}b_{n}}^{c_{n}d_{n}}.
\end{equation}
where the generalized Kronecker delta function is antisymmetric in both  series of indices.  The action of pure Lovelock gravity is written as follows
\begin{equation}
S=\int\sqrt{-g}d^dx\left( \alpha_0\mathcal{L}_0+\alpha_n \mathcal{L}_n-F_{\mu\nu}F^{\mu\nu}\right).
\end{equation}
For spherically symmetric solution, the metric can be written as follows \cite{cai3019, cai309, cai30}
\begin{equation}
	ds^2=-f\left(r\right)dt^2+{f\left(r\right)}^{-1}\ dr^2+r^2d\Omega_{d-2}^2,
\end{equation}
where 
\begin{equation}
	f\left(r\right)=1\pm r^2 \left(\frac{16 \pi  G M }{\Omega_{d - 2}\hat{\alpha}_n (d-2)r^{d-1}}-\frac{32 \pi ^2 Q^2 }{\Omega_{d - 2}^2\hat{\alpha}_n (d-3) (d-2)r^{2(d-2)}}-\frac{1}{\hat{\alpha}_n l^2}\right)^{1/n},\label{eq:laps}
\end{equation}
where $\Omega_{d-2}=2\pi^{(d-1)/2}/\Gamma[{(d-1)/2}]$ is the volume of $(d-2)$-dimension unite sphere,  ${\hat{\alpha}}_0=-1/l^2$  and $l$ is anti-de Sitter (AdS) length. The $Q$ and $M$ denote the electric charge and mass of the black hole, respectively. The positive sign  may be used when $n$ is even, while for all dimensions $ f(r) $ takes the negative sign in  Eq.\eqref{eq:laps} where $n=(d-2)/2$.  Instead of continuing
 with coupling constants $\alpha_{n}$, we utilize re-scaled form  $\hat{\alpha}_n$ and substitute them with the following relations
\begin{equation}
\begin{split}
{\hat{\alpha}}_0=\frac{\alpha_0}{\left(d-1\right)\left(d-2\right)}\ \ \ \ \ \ ,\ \ \  {\ \hat{\alpha}}_1=1\\   {\hat{\alpha}}_n=\alpha_n\prod_{i=3}^{2m}\left(d-i\right).\ \ \ \ \ \ \ \text{for} \ \ \ \ n>1
\end{split}
\end{equation}
 Using Eq.\eqref{eq:laps} and set the condition $f(r_+)=0$  where $r_+$ denotes event horizon radius,  we  arrive at 
\begin{equation}
M=\frac{\Omega_{d - 2} (d-2) r_+^{d-1}}{16 \pi  G }\left( \frac{\hat{\alpha}_n}{ r_+^{2 n}}+\frac{1}{l^2}+\frac{32 \pi^2  Q^2 r_+^{4-2d}}{\Omega^2_{d - 2}(d-2) (d-3)}\right) .\label{eq:masspure}
\end{equation}
The Hawking temperature is obtained by surface gravity as fallows: 
\begin{equation}
	T=\frac{\kappa}{2\pi}=\frac{f^\prime\left(r\right)|_{r=r_+}}{4\pi},\label{eq:temp}
\end{equation}
where $\kappa$ denotes surface gravity of black holes. By using Eq.\eqref{eq:laps} in Eq.\eqref{eq:temp} temperature has the   following form:
\begin{equation}
	T=\frac{(d-2 n-1)}{4\pi n r_+}+\frac{(d-1)}{4\pi n  \hat{\alpha}_n l^2}r_+^{2n-1}-\frac{8 \pi l^2 Q^2}{ (d-2)\Omega_{d - 2}^2 n \hat{\alpha}_n} r^{3-2 d+2 n}.\label{eq:T}
\end{equation}
Entropy, mass and temperature are related by the first law of thermodynamic where the black holes satisfy it as a thermodynamic system, too. The first law is written in the form of $dM=TdS$, therefore we obtain
\begin{equation}
S=\int\frac{1}{T}dM=\int_{0}^{r_+}{\frac{1}{T}\frac{\partial M}{\partial r_+}dr_+},\label{eq:S}
\end{equation}
using Eqs.\eqref{eq:masspure} and \eqref{eq:T} we may write entropy as follows
\begin{equation}
S=\frac{\left(d-2\right)\Omega_{d-2}{{\hat{\alpha}}_nn\ {r_+}^{d-2n}}}{4G\left(d-2n\right)}.\label{eq:SS}
\end{equation}
The thermodynamic geometry of black holes will be discussed in the next section. We obtain thermodynamic scalar curvature and thermodynamic extrinsic curvature by applying the associated entropy and thermodynamics potential relations.
\section{THERMODYNAMIC GEOMETRY OF CHARGED BLACK HOLES IN PURE LOVELOCK GRAVITY THEORY}\label{sec:pureg}
\begin{figure}[t]
	\centering
	\begin{subfigure}{.4\textwidth}
		\centering
		\includegraphics[width=1\textwidth]{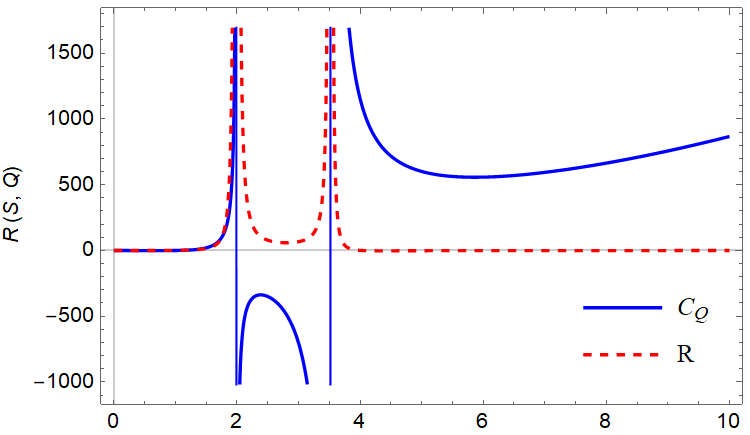}
		\caption{}
	\end{subfigure}%
	\begin{subfigure}{.4\textwidth}
		\centering
		\includegraphics[width=1\textwidth]{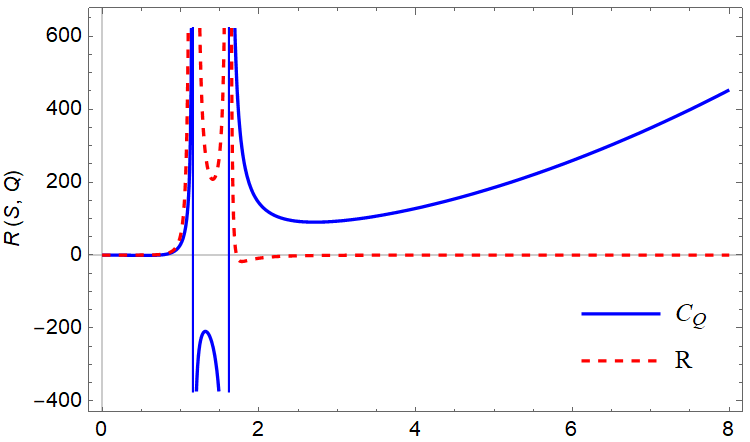}
		\caption{}
	\end{subfigure}
	\begin{subfigure}{.4\textwidth}
		\centering
		\includegraphics[width=1\textwidth]{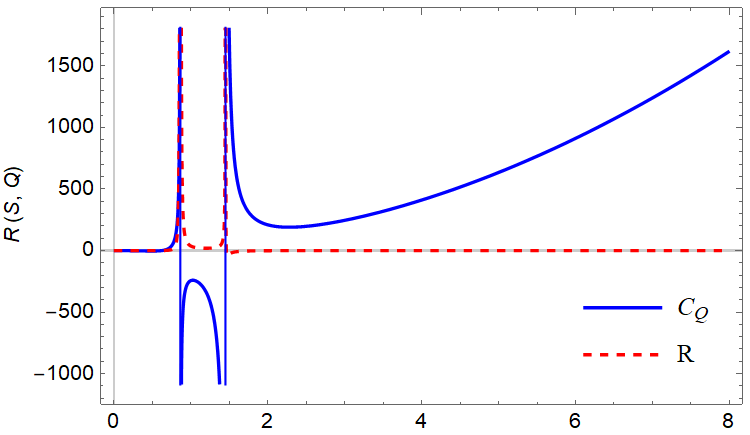}
		\caption{}
	\end{subfigure}%
	\begin{subfigure}{.4\textwidth}
		\centering
		\includegraphics[width=1\textwidth]{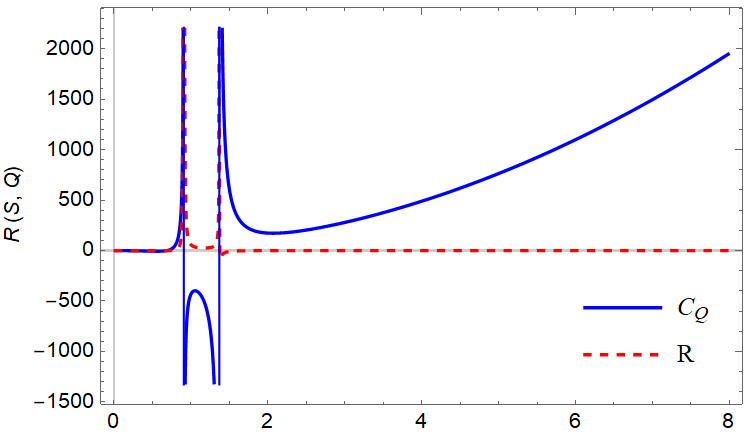}
		\caption{}
	\end{subfigure}
	\caption{\normalfont Graphs of the thermodynamic scalar curvature $R$ (dotted-dashed red line) of the new metric and the specific heat capacity $C_Q$ (solid blue line) as a function of horizon radius $r_+$. For pure Lovelock black holes in different dimensions (a) $d=4$, (b) $d=6$, (c) $d=8$, (d) $d=10$. For mentioned dimensions, we set $l=4,\ 18,\ 20,\ 30$ and $\hat{\alpha}_1=1,\ \hat{\alpha}_2=0.4,\ \hat{\alpha}_3=0.8,\ \hat{\alpha}_4=0.9$, respectively. In all diagrams we have considered $Q=1$. $G=1$ in natural unites.}
	\label{fig:wtr}
\end{figure}

We study thermodynamic geometry of charged black holes in pure Lovelock gravity. We introduce a metric that can be used to obtain some information about thermodynamic and also interaction between microstates of related black holes. It is known that there is not always a correspondence between heat capacity critical points and singularities of thermodynamic Ricci scalar in the Ruppeiner formulation of thermodynamic geometry for black holes  \cite{zamani,fail1,fail2}. Therefore, we use a modern version of the metric  for thermodynamic geometry  \cite{correspon,geoth} as follows
\begin{equation}
dl^2=\eta_i^j\frac{\partial^2\Xi}{\partial X^j\partial X^k}{{dX}}^i{{dX}}^k. \label{eq:ntg}
\end{equation}
where $\eta_{i}^{j}$ denotes the metric as $(-1,1,...,1,1)$ and $\Xi$ corresponds to the specific thermodynamic potential. Note that here we have used a conformal transformation to eliminate the $1/T$ written in \cite{correspon,geoth} since it causes an additional singular point which is related to $T=0$ that is non-physical according to third law of thermodynamic.
By using  $X^i=(S,Q)$  which shows entropy and electric charge of black holes, we study thermodynamic properties and phase transition points of black holes. In  equilibrium state space, according to the first law of thermodynamic $(dM=TdS+\Phi dQ)$, using the mass relation in Eq.\eqref{eq:masspure} and the entropy in Eq.\eqref{eq:SS} we may obtain temperature $ T $, the specific heat capacity $C_Q$, and the electric potential $\Phi  $, as follows
\begin{flalign}
&\begin{aligned}
&T=\left(\frac{\partial M}{\partial S}\right)_Q=\frac{(d-2 n-1)}{4\pi n r_+}+\frac{(d-1)}{4\pi n  \hat{\alpha}_n l^2}r_+^{2n-1}-\frac{8 \pi l^2 Q^2}{ (d-2)\Omega_{d - 2}^2 n \hat{\alpha}_n} r_+^{3-2 d+2 n},\label{eq:puret}
\end{aligned}&&
\end{flalign}
\begin{flalign}
&\begin{aligned}
&\mathrm{\Phi}=\left(\frac{\partial M}{\partial Q}\right)_S=\frac{4 \pi  Q r_+^{3-d}}{(d-3)\Omega_{d - 2}  G},\label{eq:purephi}
\end{aligned}&&
\end{flalign}
\begin{flalign}
&\begin{aligned}
&C_Q={T{\left(\frac{\partial S}{\partial T}\right)}}_Q\\ & =-\frac{\hat{\alpha}_n \Omega_{d - 2} (d-2) n r_+^{d-2 n} \left(l^2 \left(\hat{\alpha}_n \Omega_{d - 2}^2 (d-2) (d-2 n-1) r_+^{2 d-2 n}-32 \pi ^2 Q^2 r_+^4\right)+\Omega_{d - 2}^2 (d-2) (d-1) r_+^{2 d}\right)}{4 G \left(l^2 \left(\hat{\alpha}_n \Omega_{d - 2}^2 (d-2) (d-2 n-1) r_+^{2 d-2 n}+32 \pi ^2 Q^2 r_+^4 (-2 d+2 n+3)\right)-\Omega_{d - 2}^2 (d-2) (d-1) (2 n-1) r_+^{2 d}\right)}.
\end{aligned}&&
\end{flalign}
We have used the Poisson bracket method for our calculations (Appendix A).
The  roots of the heat capacity's denominator, denote phase transition points. We calculate the associated Ricci scalar and obtain singular points. To do this, by choosing the thermodynamic potential  $\Xi=M(S,Q)$  and coordinates as $X^i=(S,Q)$   in Eq.\eqref{eq:ntg}, we have
\begin{equation}
dl^2=-\frac{\partial^2M}{\partial S^2}dS^2+\frac{\partial^2M}{\partial Q^2}dQ^2.\label{eq:PG}
\end{equation}
Therefore, the  $g_{ab}$ gets the  following form
\begin{equation}
g_{ab}=\left(\begin{matrix}-\left(\frac{\partial^2M}{\partial S^2}\right)_Q&0\\0&\left(\frac{\partial^2M}{\partial Q^2}\right)_S\end{matrix}\right)=\left(\begin{matrix}-\left(\frac{\partial T}{\partial S}\right)_Q&0\\0&\left(\frac{\partial\Phi}{\partial Q}\right)_S\end{matrix}\right).
\end{equation}
The metric  elements can be written as below
\small
\begin{equation*}
g_{SS}=\frac{r_+^{-3 d+2 n-1} \left(\hat{\alpha}_n \Omega_{d - 2}^2 (d-2) G l^2 (d-2 n-1) r_+^{2 d}+G r_+^{2 n} \left(32 \pi ^2 l^2 Q^2 r_+^4 (-2 d+2 n+3)-\Omega_{d - 2}^2 (d-2) (d-1) (2 n-1) r_+^{2 d}\right)\right)}{\pi  \hat{\alpha}_n^2 \Omega_{d - 2}^3 (d-2)^2 l^2 n^2},
\end{equation*}
\normalsize
\begin{flalign}
&\begin{aligned}
g_{QQ}=\frac{4 \pi  r_+^{3-d}}{ (d-3)\Omega_{d - 2} G}.
\end{aligned}&&
\end{flalign}
Now, by  using the above metric elements, and the following formula for the Ricci scalar 
\begin{equation}
R=\frac{2\left|\begin{matrix}g_{SS}&g_{QQ}&g_{SQ}\\g_{SS,S}&g_{QQ,S}&g_{SQ,S}\\g_{SS,Q}&g_{QQ,Q}&g_{SQ,Q}\\\end{matrix}\right|}{\left|\begin{matrix}g_{SS}&g_{SQ}\\g_{SQ}&g_{QQ}\\\end{matrix}\right|^2},\label{eq:ricci}
\end{equation}
the thermodynamic Ricci scalar may be written as follows
\begin{equation}
R=-\frac{16 \pi   (d-3) (d-2)\Omega_{d - 2}^3 G l^2 (2 d-2 n-3) r_+^{3 d+2 n+1} \left(\hat{\alpha}_n l^2 (2 n-d+1)+(d-1) (2 n-1) r_+^{2 n}\right)}{\left(\hat{\alpha}_n \Omega_{d - 2}^2 (d-2) l^2 (d-2 n-1) r_+^{2 d}-r_+^{2 n} \left(\Omega_{d - 2}^2 \left(d^2-3 d+2\right) (2 n-1) r_+^{2 d}+32 \pi ^2 l^2 Q^2 r_+^4 (2 d-2 n-3)\right)\right)^2}.\label{eq:purericci}
\end{equation}
We study the pure Lovelock theory in different dimensions ($d=4,\ d=6,\ d=8,\ d=10$). The thermodynamic Ricci scalar and the specific heat capacity diagrams are depicted in Fig. \ref{fig:wtr}. In order to determine the phase transition points of the specific heat capacity, we only need  to obtain roots of Eq.\eqref{eq:purericci}.  Fig.\ref{fig:wtr} shows divergences of the thermodynamic scalar curvature that are corresponded to the phase transition points. In those regions where $C_Q$ is negative, we expect that the system to be unstable. For other regions that $C_Q$ is positive we expect stability of the system. Thermodynamic metric in Eq.\eqref{eq:PG} provides  some important information about the microstructure  of black holes. We may find some parts in parameter space with attractive or repulsive interaction between microstates. The positive values of thermodynamic Ricci scalar $(R>0)$ are corresponded to fermionic behavior of microstates which implies repulsive interactions while the negative values $(R<0)$ are related to bosonic behavior of microstates which shows attractive interaction \cite{anyon}. 
 We have depicted thermodynamic curvature $R$, and specific heat $C_Q$  as a function of horizon radius in Fig.\ref{fig:wtr}. It is seen that the thermodynamic curvature of the metric is exactly similar to specific heat capacity $C_Q$ at transition points. As  expected, there is repulsive interaction and fermionic behavior in the region of $R>0$. 
 \begin{figure}[t]
	\centering
	\begin{subfigure}{.4\textwidth}
		\centering
		\includegraphics[width=1\textwidth]{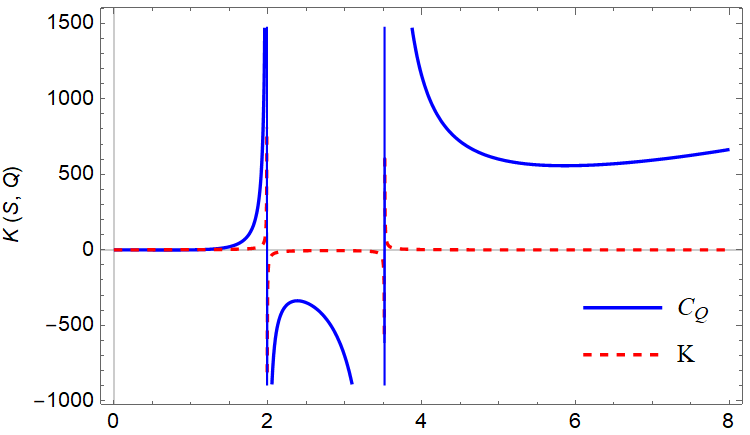}
		\caption{}
	\end{subfigure}%
	\begin{subfigure}{.4\textwidth}
		\centering
		\includegraphics[width=1\textwidth]{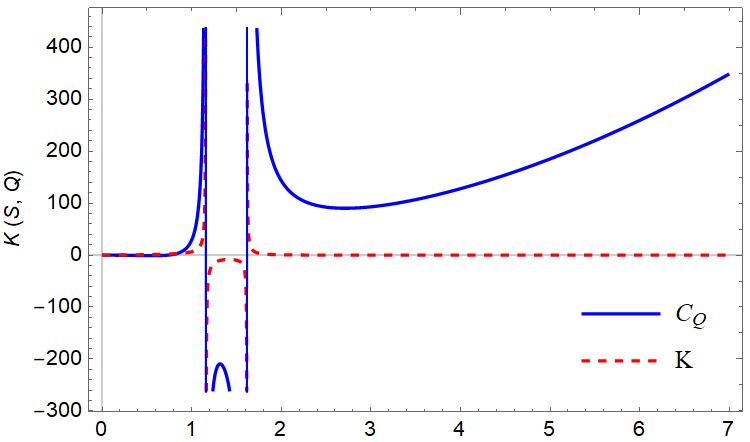}
		\caption{}
	\end{subfigure}
	\begin{subfigure}{.4\textwidth}
		\centering
		\includegraphics[width=1\textwidth]{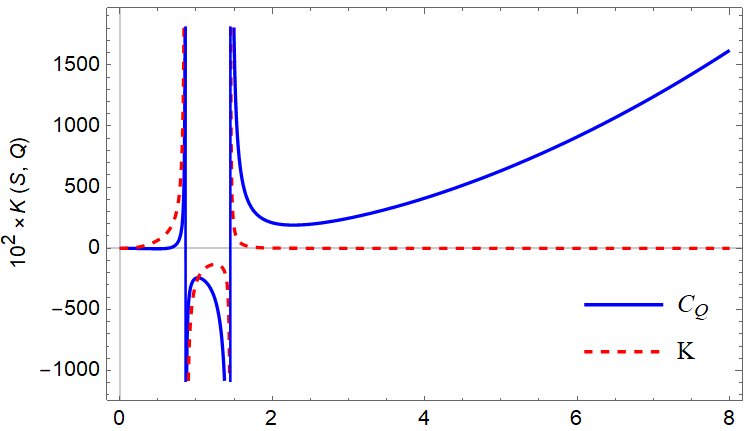}
		\caption{}
	\end{subfigure}%
	\begin{subfigure}{.4\textwidth}
		\centering
		\includegraphics[width=1\textwidth]{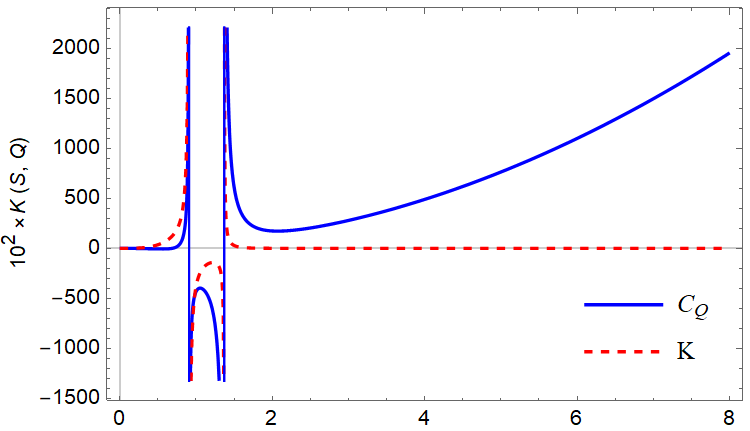}
		\caption{}
	\end{subfigure}
	\caption{ \normalfont Graphs of the thermodynamic extrinsic curvature $K$ (dashed  red curve) and the specific heat capacity $C_Q$ (solid blue curve) as a function of horizon radius $r_+$ for pure Lovelock black holes with (a) $d=4$, (b) $d=6$, (c) $d=8$, (d) $d=10$. For mentioned dimensions, we set $l=7,\ 18,\ 20,\ 30$ and $\hat{\alpha}_1=1,\ \hat{\alpha}_2=0.4,\  \hat{\alpha}=0.8,\ \hat{\alpha}_4=0.9$, respectively. In all diagrams we have considered $Q=1$. $G=1$ in natural unites.}
	\label{fig:purex}	
\end{figure}
According to Fig.\ref{fig:wtr}, it is clear that the sign of $R$ is not the same as $C_Q$ in all regions  therefore it does not give us the precise information about the stability or instability of the  thermodynamic system.  In order to do that, we need another geometrical quantity called extrinsic curvature \cite{extrin}.
 
Now, we review some aspects of the extrinsic curvature. For hypersurface $\Sigma$ which is embedded in thermodynamic  manifold $\mathcal{M}$, the extrinsic curvature is given by 
\begin{equation}
K=\mathrm{\nabla}_\mathrm{\mu}n^\mu=\frac{1}{\sqrt{g}}\partial_\mu\left(\sqrt{ g}n^\mu\right),\label{eq:ext}
\end{equation}
\begin{figure}[t]
	\centering
	\begin{subfigure}{.4\textwidth}
		\centering
		\includegraphics[width=1\textwidth]{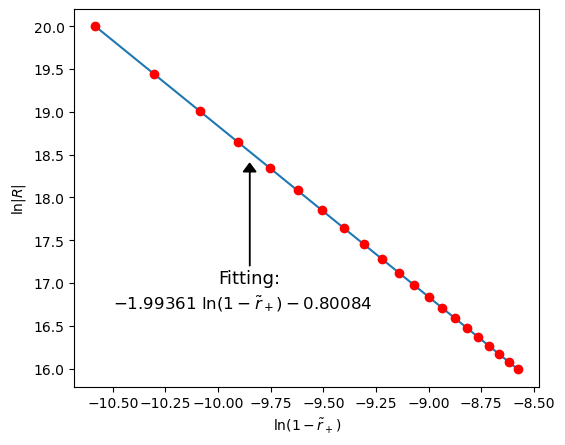}
		\caption{}
	\end{subfigure}%
	\begin{subfigure}{.4\textwidth}
		\centering
		\includegraphics[width=1\textwidth]{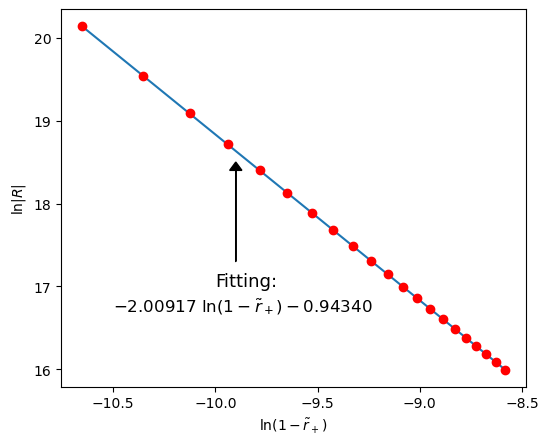}
		\caption{}
	\end{subfigure}
	\begin{subfigure}{.4\textwidth}
		\centering
		\includegraphics[width=1\textwidth]{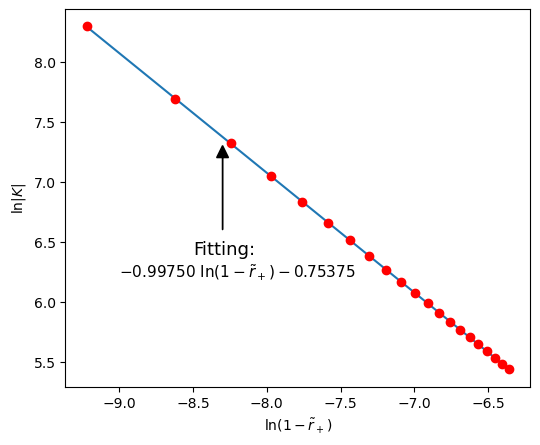}
		\caption{}
	\end{subfigure}%
	\begin{subfigure}{.4\textwidth}
		\centering
		\includegraphics[width=1\textwidth]{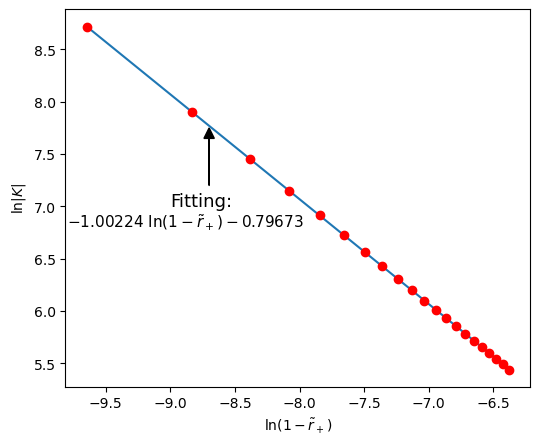}
		\caption{}
	\end{subfigure}
	\caption{\normalfont Logarithm of  thermodynamic scalar curvatures as a function of $\ln(1-\tilde{r}_+)$  in  6-dimensions.  The numerical data are shown by  red markers and the solid blue line is plotted by the  fitting formulas. (a) $\ln \left| R \right| \ $ diagram,  along small black hole coexistence curve, the slope is $-1.99361$. (b) $ \ln \left| R \right| \ $ diagram, along large black hole coexistence curve, the slope is $-2.00917$.  (c) $ \ln \left| K \right| \ $ diagram, along small black hole coexistence curve, the slope is $-0.99750$. (b) $ \ln \left| K \right| \ $  diagram, along large black hole coexistence curve, the slope is $-1.00224$.}
	\label{fig:coex1}
\end{figure}
where $n_\mu$ is a normal vector to hypersurface $\Sigma$ is written as follows
\begin{equation}
n_\mu=\frac{\partial_\mu\mathcal{H}}{\sqrt{g^{\mu\nu}\partial_\mu\mathcal{H}\partial_\nu\mathcal{H}}}.\label{eq:vn}
\end{equation}
The hypersurface $\Sigma$ is defined by $\mathcal{H}(X^\alpha)=0$, and $X^\alpha$ represents coordinates of the manifold $\mathcal{M}$. The extrinsic curvature is a novel tool to study thermodynamic properties of physical systems \cite{extrin}. To  investigate  $C_Q$, we stick to a constant $Q$ hypersurface where $\mathcal{H}=Q-\mathbb{C}=0$ and $\mathbb{C}$ is a constant, and therefore normal vector $n_Q$ reads
\begin{equation}
n_Q=\frac{1}{\sqrt{g^{QQ}}}=\sqrt{\frac{4 \pi  r_+^{3-d}}{ (d-3)\Omega_{d - 2} G}},\label{eq:nv}
\end{equation}
As a consequence, by using Eq.\eqref{eq:nv} and  square root of determinant of the metric ($\sqrt{g}$), extrinsic curvature  gets the following form $(K=\frac{1}{\sqrt{g}}\partial_Q(\sqrt{g}n^Q))$
\begin{equation}
K_{\text{PL}}=\frac{16 \pi ^{3/2} G l^2 Q (-2 d+2 n+3) r_+^{2 n+4} \sqrt{ (d-3)\Omega_{d - 2} G r_+^{d-3}}}{\hat{\alpha}_n (d-2)\Omega_{d - 2} ^2  G l^2 (d-2 n-1) r_+^{2 d}+G r_+^{2 n} \left(32 \pi ^2 l^2 Q^2 r_+^4 (-2 d+2 n+3)-\Omega_{d - 2} ^2 (d-2) (d-1) (2 n-1) r_+^{2 d}\right)}\label{eq:extrinsic}.
\end{equation}	
It is seen in Fig.\ref{fig:purex} that the thermodynamic extrinsic curvature provides  more useful  information than the thermodynamic Ricci scalar. Not only it has the same divergence points as the phase transitions in $C_Q$, but also contains correct information about stability and instability of the system around phase transition points. It is  seen in Fig.\ref{fig:purex}, the black hole  is unstable  around phase transition points, however, it is stable in other regions. We explore the critical behavior of the thermodynamic Ricci scalar and extrinsic curvature around the critical points where they diverge. A hypothesis on dimensional grounds states that $ R\sim \xi^d$ where $\xi$ is correlation length which is the only scale exists near the critical regions \cite{Rupp,riccialpha,prof2}. Using  $R\sim \xi^d$ and the hyperscaling, $\nu d=2-\alpha$ where $\nu$ and $\alpha$ are critical exponents, we expect  the following relation
 \begin{equation}
 	R \sim t^{\alpha-2},\label{eq:sim}
 \end{equation}
where $t=1-T/T_C$ and $T_C$ denotes temperature value at the critical point. Therefore, the scaling behavior of $R$ in Eq.\eqref{eq:sim} provides a possibility of deriving the  critical exponent $\alpha$.  The above relation is correct for $\alpha>0$. It was found that there are two standard scaling forms as $R\sim t^{\alpha-2}$ and $R\sim t^{\alpha-1}$ for $\alpha > 0$ and $\alpha<0$, respectively \cite{riccialpha,fla1,fla2}. Moreover,  for $\alpha=0$ in 2-dimensional kagome Ising model under the presence of an external field, thermodynamic Ricci scalar near the critical region behaves as $R\sim t^{\alpha-1}$ \cite{prof2}. The scaling behavior of thermodynamic curvature for $\alpha=0$ in four-dimensional spherical and three-dimensional Van-der Waals model  is $R\sim t^{\alpha-2}$. Therefore, for $\alpha=0$ thermodynamic Ricci scalar has a dimension dependent scaling behavior. We have found that the scaling behavior of the Ricci scalar curvature for our model is $R\sim t^{\alpha-2}$ where   $\alpha=0$. Now, we investigate the critical behavior of the scalar curvatures for pure Lovelock black holes close to the critical point. This provides some universal behaviors of pure Lovelock black holes. Here, we investigate six-dimensional pure Lovelock black holes. By setting the values $l=18, \hat{\alpha}_2=0.4$ and $Q=1$ in the both thermodynamic Ricci scalar and the extrinsic curvature relations in Eqs.\eqref{eq:purericci} and \eqref{eq:extrinsic} for six-dimensional pure Lovelock black holes, we have
 \begin{equation}
 	R\approx -\frac{1.5\times 10^6 r_+^7 \left(r_+^4-9\right)}{\left(1200+\pi ^4 r_+^8-842 r_+^4\right)^2}, \label{eq:ricci04}
 \end{equation}
 \begin{equation}
 	K\approx \frac{31 \sqrt{r_+^3}}{r^8-9 r_+^4+12}. \label{eq:ext4d}
 \end{equation}  
 Near the critical points, thermodynamic Ricci scalar and thermodynamic extrinsic curvature, respectively, can be written as  
 \begin{equation}
 	\begin{aligned}
 		&	R\propto \left( 1- \tilde{r}_+\right) ^{-a} \; \ \ \ \ \ \ \ \ \text{or} \ \ \ \ \ \ \ \ \ln|R|=a\ln\left(1-\tilde{r}_+ \right)+b,\\ 
 		&   K\propto -\left( 1- \tilde{r}_+\right) ^{-c} \; \ \ \ \ \ \text{or} \ \ \ \ \ \ \ \ln|K|=-c\ln\left(1-\tilde{r}_+ \right)+d,
 		\label{eq:lnr}
 	\end{aligned}
 \end{equation}
  \begin{table}[t]
	\begin{center}
		\caption {\normalfont Critical coefficients of fitting formulas for small and large black hole where $n$ shows the order of the gravity. }\label{tab:1}
		\begin{tabular}{|c|c|c|c|c||c|c||c|c||c|}
			\toprule
         	& $n$	& $dim$  & $\hat{\alpha}_n$ & $l$ & $a \ (R)$ & $-b \ (R) $ & $c \ (K)$ & $-d ( K)$ &  $\alpha$   \\
			\midrule
			Small BH & 1	&	4  & 1 & 7 &   1.99772  & 2.50992 & 0.99934  & 2.25088 &	0    \\
			Large BH &     &     & & & 2.00172  & 2.91116 &  1.00050  &  2.27688  & 0 \\ 
			\hline
			Small BH & 2	&	6   & 0.4 & 18 &  1.99361  & 2.21684 & 0.99750  & 2.12496 &	0    \\
			Large BH &     &      & & & 2.00917  & 2.56872 &  1.00224  & 2.21829  & 0 \\ 
			\hline
			Small BH & 3	& 8 & 0.8 & 20  & 2.00625 & 2.74091  & 1.00028  &2.28750 & 0   \\
			Large BH &     &      & &  & 1.99683 & 2.52757 &  0.99990  &  2.27927  & 0 \\ 
			\hline
			Small BH & 4	&	10   & 0.9 & 30 &  2.00754  & 3.73042 & 1.00028  &2.65384 &	0    \\
			Large BH &     &      & & & 1.99503  & 3.35170 &  0.99981  &  2.64320  & 0 \\ 
			\bottomrule
		\end{tabular}
	\end{center}
\end{table}

\begin{table}[t]
	\begin{center}
		\caption {\normalfont Critical amplitude of thermodynamic curvatures near the critical point, evaluated by numerical method. }\label{tab:2}
		\begin{tabular}{|c|c|c|c|}
			\toprule
			$n$	& $dim$   & $R(1-\tilde{r_+})^2$ & $K(1-\tilde{r_+})$ \\
			\midrule
			1	&	4   &  0.0772630  & 0.103955  \\
			\hline
			2	&	6   &  0.0913753  & 0.113992   \\
			\hline
			3 &   8   & 0.0717735  & 0.101939 \\ 
			\hline
			4   &   10   &  0.0289826  & 0.070756  \\ 
			\bottomrule
		\end{tabular}
	\end{center}
\end{table}
where we have used the reduced parameter  $\tilde{r}_+=r_+/r_{c_+}$ and $r_{c_+}$ is the value of horizon radius at a critical point where $R$ and $K$ diverge.  Note that in Fig.\ref{fig:wtr}, it is found that we have two critical points in which $R$ diverges. Here, we investigate the scaling behavior of scalar curvatures near one of the critical points in which representing the universal feature. By calculating the critical values of the horizon radius $\tilde{r}_{c_+}$ in Eqs. \eqref{eq:ricci04} and \eqref{eq:ext4d} where thermodynamic Ricci scalar and extrinsic curvature diverge, one can fit the formula near critical points by  numerical method and gets the following results
\begin{align}
 &\ln|R|=-1.99361\ln\left(1-\tilde{r}_+ \right) -2.21684, \ \ \ \text{Small balck hole} \label{eq:lnr1} \\ 
 &\ln|R|=-2.00917 \ln\left( 1-\tilde{r}_+ \right) -2.56872. \ \ \ \text{Large black hole} \label{eq:lnr2}
\end{align}
 Also, extrinsic curvature in Eq.\eqref{eq:ext4d} yields
\begin{align}
 &\ln|K|=-0.99750 \ln\left( 1-\tilde{r}_+ \right) -2.12496, \ \ \ \text{Small black hole}\\
 &\ln|K|=-1.00224 \ln\left( 1-\tilde{r}_+ \right) -2.21829. \ \ \ \text{Large black hole}
\end{align}
The numerical results for various dimensions are reported in Table \ref{tab:1}. Using Eqs.\eqref{eq:lnr1} and \eqref{eq:lnr2} we can compute the critical amplitude by using the constants $b$ and $d$ as follows
\begin{equation}
	R(1-\tilde{r}_+)^2=e^{-\frac{2.21684+2.56872}{2}}\approx 0.08,
\end{equation}
and for extrinsic curvature we have
\begin{equation}
	K(1-\tilde{r}_+)=e^{-\frac{2.12496+2.21829}{2}} \approx 0.12.
\end{equation}
The numerical results of the critical amplitudes  are  given in Table \ref{tab:2}. 
We have shown the fitting results for $ d=6 $ in Fig.\ref{fig:coex1}. The red markers show numerical data and the solid blue line indicates the fitting formulas. So, by considering the numerical method errors, it is clear that $a=2$, $c=1$ and therefore $ \alpha $ can be found as 
 \begin{equation}
 	\alpha=0.
 \end{equation}
Therefore, the scaling behavior is similar to  Van der Waals fluid in $d=3$ and the spherical model in $d=4$. We avoid  writing computation of other dimensions here. See the results of fitting formula by the numerical method in Table \ref{tab:1}.

\section{EHRENFEST APPROACH TO PHASE TRANSITION }\label{sec:ehrenfest}
It is explained that the phase transition points of the specific heat capacity have a crucial role to find out the thermodynamic behavior of black holes. Ehrenfest's approach can be utilized to better understand critical behavior of different kinds of thermodynamic system \cite{ehren1,ehren2,ehren3}. In the Ehrenfest's approach, the phase transition order is considered as the  lowest order of  Gibbs free energy differential in which we have discontinuities through the phase transition. Based on the first law of thermodynamic $dM=TdS+\Phi dQ$, the term $\Phi dQ$, is similar to $-PdV$ in typical thermodynamic, representing the work term which is essential to obtain the Ehrenfest's equations. The free energy can be defined as $G=M-TS$, in the canonical ensemble. Considering the first law, then we have
\begin{equation}
dG=-SdT +\Phi dQ.
\end{equation}
Also, if we consider entropy as $S=S(T,Q)$, then we will have
\begin{equation}
	dS= \left( \frac{\partial S}{\partial T}\right)_Q dT +  \left( \frac{\partial S}{\partial Q}\right)_T dQ,
\end{equation}
which can be written as follows
\begin{equation}
	dS=\frac{C_Q}{T}dT + \Phi \bar{\alpha} dQ \label{eq:dS1},
\end{equation}
where $C_Q=T(\partial S/\partial T)_Q$ and $\bar{\alpha}=\frac{1}{\Phi}(\partial S/\partial Q)_T$. We have $S_1 = S_2$ and $\Phi_1=\Phi_2$ where subscripts 1 and 2 indicate the phase before and after the second order phase transition . As a result, $dS_1 = dS_2$, and by using \eqref{eq:dS1} we have the following first Ehrenfest's equation
\begin{equation}
-\left( \frac{\partial Q}{\partial T}\right)_S=\frac{1}{\Phi T}\frac{\left( C_Q\right)_2 -\left( C_Q\right)_1 }{\bar{\alpha}_2-\bar{\alpha}_1},\label{eq:eh0}
\end{equation}
A similar procedure can be used for $\Phi$ as follows
\begin{equation}
d\Phi = - \Phi \bar{\alpha} dT + \Phi \kappa dQ,
\end{equation}
where $\kappa = \frac{1}{\Phi} (\partial \Phi / \partial Q)_T$. Also, form the $d\Phi_1 = d\Phi_2$ we have the second Ehrenfest's equation as below
\begin{equation}
\left( \frac{\partial Q}{\partial T}\right)_\Phi=\frac{\bar{\alpha}_2-\bar{\alpha}_1}{\kappa_2-\kappa_1},\label{eq:eh00}
\end{equation}
Assuming the grand canonical ensemble, $Q$ is not a constant, however, its conjugate $\Phi$, is fixed. Choosing free energy as $\tilde{G} = M -TS -\Phi Q $, we obtain the Ehrenfest's equations as follows
\begin{equation}
-\left( \frac{\partial \Phi}{\partial T}\right)_S=\frac{1}{QT}\frac{\left( C_\Phi\right)_2 -\left( C_\Phi\right)_1 }{\bar{\alpha}_2-\bar{\alpha}_1},\label{eq:eh1}
\end{equation}
\begin{equation}
-\left( \frac{\partial \Phi}{\partial T}\right)_Q=\frac{\bar{\alpha}_2-\bar{\alpha}_1}{\kappa_2-\kappa_1},\label{eq:eh2}
\end{equation}
where $C_\Phi=T(\partial S/\partial T)_\Phi$. $\kappa$ and $\bar{\alpha}$ indicate the  isothermal compressibility and volume expansion coefficient respectively, and can be written as follows
\begin{equation}
\bar{\alpha} = -\frac{1}{Q}\left(\frac{\partial S}{\partial \Phi} \right)_T =\frac{1}{Q}\left(\frac{\partial Q}{\partial T} \right)_\Phi, \ \ \ \ \ \ \ \ \ \ \ \ \ \ \ \  \kappa=\frac{1}{Q}\left(\frac{\partial Q}{\partial \Phi} \right)_T. \label{eq:alphakappa}
\end{equation}
The generalized form of the Ehrenfest's equations for different kinds of black holes have been investigated  in \cite{ehren3}. We may examine the validity of the Ehrenfest equations for pure Lovelock black hole. At the critical point, the left hand side  of Eq.\eqref{eq:eh1} may be written as
\begin{equation}
\begin{aligned}
&-\left( \frac{\partial \Phi}{\partial T}\right)_{S=S_c}=-\left( \frac{\partial \Phi}{\partial T}\right)_{r_+=r_{c_+}}\\
& \ \ \ \ \ \ \ \ \ \ \ \ \ \ \ \ \ \  \ =-\left(\frac{\partial \Phi}{\partial Q}\right)_{r_+=r_{c_+}}\left(\frac{\partial T}{\partial Q}\right)^{-1}_{r_+=r_{c_+}},
\end{aligned}
\end{equation}   
where $c$ indicates the critical value of the associated quantity. By using the chain rule and derivative of  Eq.\eqref{eq:puret} and Eq.\eqref{eq:purephi} with respect to $Q$ at critical values, the left hand side of Eq.\eqref{eq:eh1} is given by
\begin{equation}
	-\left( \frac{\partial \Phi}{\partial T}\right) _{r_+=r_{c_+}}=\frac{(d-2)\Omega_{d - 2} \hat{\alpha}_n n r_{c_+}^{d-2 n}}{4 (d-3) G Q_c}.\label{eq:purelhs1}
\end{equation}
On the other hand,  using $C_\Phi=T(\partial S/\partial T)_\Phi$ and the  defination of $ \bar{\alpha} $ in Eq.\eqref{eq:alphakappa}, the right hand side of Eq.\eqref{eq:eh1} is rewritten in the following form
\begin{equation}
	\frac{\left( C_\Phi\right)_2 -\left( C_\Phi\right)_1 }{QT(\bar{\alpha}_2-\bar{\alpha}_1)}=\left( \frac{\partial S}{\partial Q}\right)_\Phi\label{eq:rhs1}
\end{equation}
then, by using  the  following identity  $\left(\frac{\partial Q}{\partial S} \right)_\Phi	\left(\frac{\partial S}{\partial \Phi} \right)_Q	\left(\frac{\partial \Phi}{\partial Q} \right)_S=-1$, the Eq.\eqref{eq:rhs1} can be written as the form below
\begin{equation}
	\frac{\left( C_\Phi\right)_2 -\left( C_\Phi\right)_1 }{QT(\bar{\alpha}_2-\bar{\alpha}_1)}=-\left( \frac{\partial \Phi}{\partial Q}
	\right)_{r_+=r_{c_+}} \left(\frac{\partial \Phi}{\partial r_+} \right)^{-1}_{Q}
	\left( \frac{dS}{dr_+}
	\right)_{r_+=r_{c_+}}. \label{eq:deltaC}
\end{equation}
Now,  we use  the relations Eq.\eqref{eq:SS} and  Eq.\eqref{eq:purephi}, respectively,  for $ S $ and   $ \Phi  $    in Eq.\eqref{eq:deltaC}. Therefore,  doing a  simple computation  the first Ehrenfest's equation may be written as follows
\begin{equation}
	\frac{\left( C_\Phi\right)_2 -\left( C_\Phi\right)_1 }{QT(\bar{\alpha}_2-\bar{\alpha}_1)}=\frac{(d-2)\Omega_{d - 2} \hat{\alpha}_n n r_{c_+}^{d-2 n}}{4 (d-3) G Q_c}.\label{eq:purerhs1}
\end{equation}
 Comparing Eq.\eqref{eq:purelhs1} and Eq.\eqref{eq:purerhs1} indicates that the first Ehrenfest's equation i.e. Eq.\eqref{eq:eh1} is valid for pure Lovelock black holes.\\
 To confirm that the second Ehrenfest equation in \eqref{eq:eh2} is correct,  we consider temperature as $T \equiv T(S,\Phi)$  therefore we have 
 \begin{equation}
   \left(\frac{\partial T}{\partial \Phi} \right)_Q =	\left(\frac{\partial T}{\partial S} \right)_\Phi	\left(\frac{\partial S}{\partial \Phi} \right)_Q+	\left(\frac{\partial T}{\partial \Phi} \right)_S . \label{eq:dtdphi}
 \end{equation}
At the critical points we have $(\partial T/\partial S)_\Phi=0$, therefore, by using  Eq.\eqref{eq:dtdphi}, the left hand side of the Eq.\eqref{eq:eh2} can be written as follows
\begin{equation}
 \left(\frac{\partial T}{\partial \Phi} \right)_{Q}
=\left( \frac{\partial T}{\partial \Phi} \right)_{S={S_c}}=\frac{(d-2)\Omega_{d - 2} \hat{\alpha}_n n r_{c_+}^{d-2 n}}{4 (d-3) G Q_c}.\label{eq:rela}
\end{equation}
Using $ \left(\frac{\partial Q}{\partial \Phi} \right)_T	\left(\frac{\partial \Phi}{\partial T} \right)_Q	\left(\frac{\partial T}{\partial Q} \right)_\Phi=-1 $, one can find the relation between  the volume expansion coefficient $\bar{\alpha}$  and the isothermal compressibility $ \kappa$ at the critical points as below 
\begin{equation}
\kappa Q_c=\left(\frac{\partial T}{\partial \Phi} \right)_Q\bar{\alpha} Q_c.
\end{equation}
Therefore, by using Eq.\eqref{eq:rela}  we can get the right hand side of  Eq.\eqref{eq:eh2} in the following form
\begin{equation}
\begin{aligned}
&\frac{\bar{\alpha}_2-\bar{\alpha}_1}{\kappa_2-\kappa_1}=-\left(\frac{\partial\Phi}{\partial T}\right)_{r_+=r_{c_+}}\\
& \ \ \ \ \ \ \ \ \ \ \ =\frac{(d-2)\Omega_{d - 2} \hat{\alpha}_n n r_{c_+}^{d-2 n}}{4 (d-3) G Q_c}.\label{eq:rhs2}
\end{aligned}
\end{equation}
We conclude  the second Ehranfest's equation is valid for pure Lovelock black holes and therefore we expect a second order phase transition. The Prigogine-Defay ratio can be also verified for this type of black holes \cite{PD1,PD2,PD3}. According to Eq.\eqref{eq:rela} and the first Ehrenfest equation we have 
\begin{equation}
\begin{aligned}
	&\left(\frac{\partial \Phi}{\partial T}\right)_{S=S_c}=\left(\frac{\partial \Phi}{\partial T} \right)_{Q}\\
	& \ \ \ \ \ \ \ \ \ \ \ \ \ \ \ \ =-\frac{\left( C_\Phi\right)_2 -\left( C_\Phi\right)_1 }{QT(\bar{\alpha}_2-\bar{\alpha}_1)}.
\end{aligned}
\end{equation} 
By using the above equation and the relation in Eq.\eqref{eq:rhs2}  the Prigogine-Defay ratio $\Pi$, is given by
\begin{equation}
	\Pi=\frac{\left( C_\Phi\right)_2 -\left( C_\Phi\right)_1 }{QT(\bar{\alpha}_2-\bar{\alpha}_1)}=\frac{\Delta C_\Phi \Delta\kappa}{T_cQ_c(\Delta\bar{\alpha})^2}=1.
\end{equation}
  Using $\Pi=1$, and the Ehrenfest's equations, we have proved that the a second order phase transition appears in thermodynamics of pure Lovelock black holes. We study the critical behavior of pure Lovelock black holes in the section.
\section{CRITICAL BEHAVIOR AND THERMODYNAMIC GEOMETRY OF PURE LOVELOCK BLACK HOLES IN THE EXTENDED PHASE SPACE}\label{sec:exten}
\begin{figure}[t]
	\centering
	\begin{subfigure}{.35\textwidth}
		\centering
		\includegraphics[width=1\textwidth]{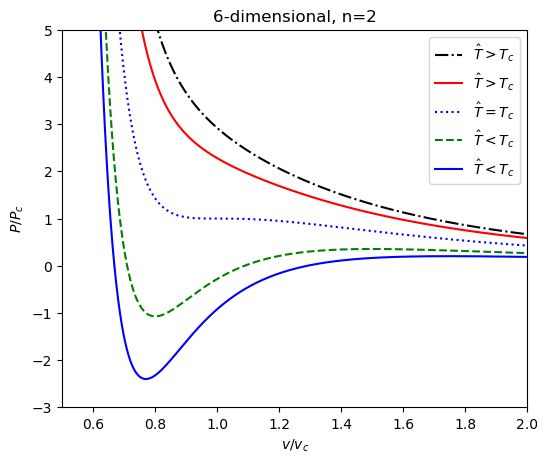}
		\caption{}
	\end{subfigure}%
	\begin{subfigure}{.35\textwidth}
		\centering
		\includegraphics[width=1\textwidth]{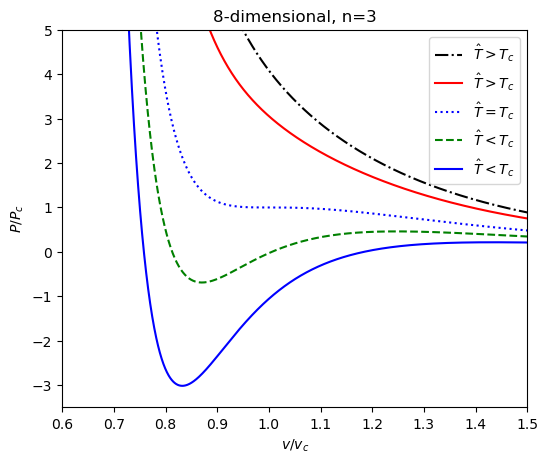}
		\caption{}
	\end{subfigure}
	\begin{subfigure}{.35\textwidth}
		\centering
		\includegraphics[width=1\textwidth]{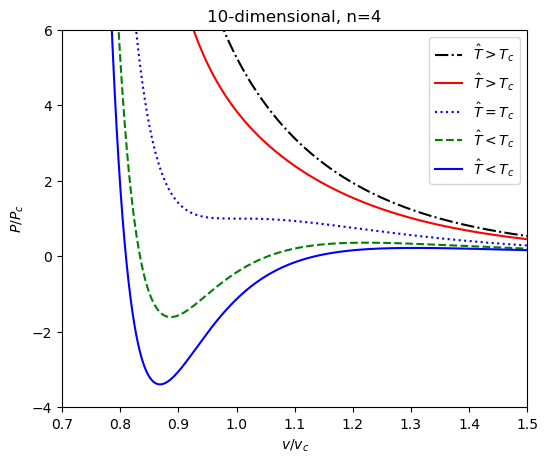}
		\caption{}
	\end{subfigure}
	\caption{\normalfont Isotherm $\hat{P}-\hat{v}$ diagrams in various dimensions $d=6, \ d=8 \ \text{and} \ d=10$, for different temperatures which are larger and smaller than the critical temperature $\hat{T}=1$. }
	\label{fig:p-v1}
\end{figure}

\begin{figure}[t]
	\centering
	\begin{subfigure}{.35\textwidth}
		\centering
		\includegraphics[width=1\textwidth]{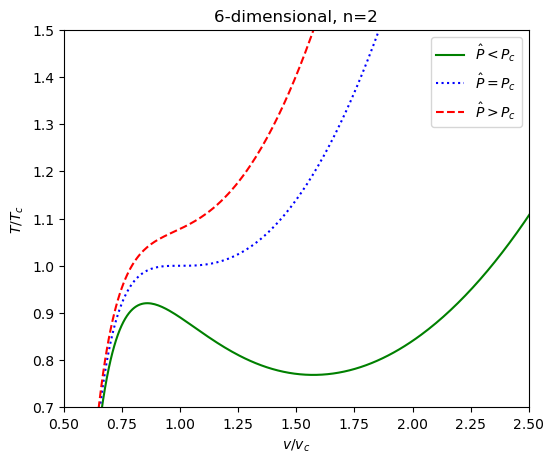}
		\caption{}
	\end{subfigure}%
	\begin{subfigure}{.35\textwidth}
		\centering
		\includegraphics[width=1\textwidth]{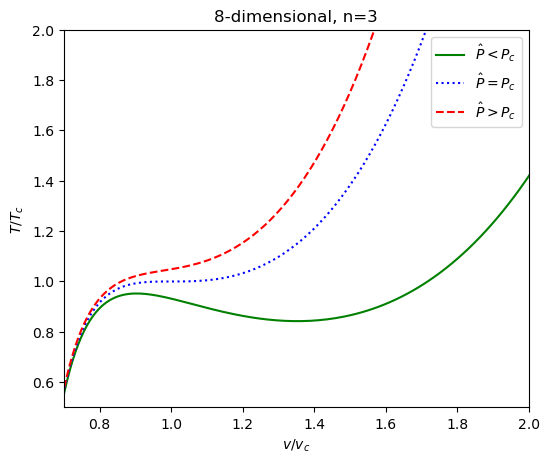}
		\caption{}
	\end{subfigure}
	\begin{subfigure}{.35\textwidth}
		\centering
		\includegraphics[width=1\textwidth]{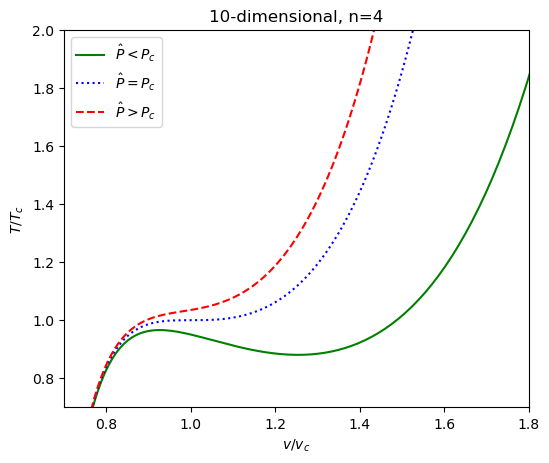}
		\caption{}
	\end{subfigure}
	\caption{\normalfont Isobaric curve $\hat{T}-\hat{v}$ diagrams in various dimensions $d=6, \ d=8 \ \text{and} \ d=10$, for different pressures which are larger and smaller than the critical value $\hat{P}=1$. }
	\label{fig:T-v1}
\end{figure}

In the following lines, we investigate thermodynamic properties and thermodynamic geometry of pure Lovelock black holes in the extended phase space. The thermodynamic pressure  is written in terms of AdS length $l$ as below 
\begin{equation}
P=\frac{(d-1)(d-2)}{16\pi l^2}.\label{eq:pres}
\end{equation}
The first law of thermodynamic for charged AdS black holes in the extended phase space can be written in the following form 
\begin{equation}
dM=TdS+VdP+\Phi dQ.\label{eq:1st}
\end{equation}
Therefore, black hole's mass might be considered as enthalpy instead of internal energy \cite{mann,kastor}. Using Eq.\eqref{eq:pres}, one can rewrite the laps function $f(r)$ in Eq.\eqref{eq:laps} with respect to thermodynamic variables as below 
\begin{equation}
	f(r)=1-r^2 \left(\frac{16 \pi  G M r^{1-d}}{\hat{\alpha}_n  (d-2)\Omega_{d - 2}}-\frac{16 \pi  P}{\hat{\alpha}_n (d-2) (d-1)}-\frac{32 \pi ^2 Q^2 r^{4-2 d}}{\hat{\alpha}_n \Omega_{d - 2}^2 (d-3) (d-2)}\right)^{1/n},\label{eq:lapsp}
\end{equation}
By using $f(r_+)=0$, the mass of black hole is given by
\begin{equation}
M=	\frac{\Omega_{d - 2} r_+^{d-1}}{(d-1)G}P+\frac{\hat{\alpha}_n(d-2)\Omega_{d - 2}r_+^{d-2n-1} }{16 \pi G}+\frac{2 \pi  Q^2 r_+^{3-d}}{(d-3)\Omega_{d - 2} G}.\label{eq:massp}
\end{equation} 
Now, according to the first law of thermodynamic in Eq.\eqref{eq:1st}, the thermodynamic volume $V$, electric potential $\Phi$ and the temperature $T$, of black hole can be written as follows 
\begin{equation}
	V=\left( \frac{\partial M}{\partial P}\right) _{S,Q}=\frac{ \Omega_{d - 2} r_+^{d-1}}{(d-1) G}=\frac{\Omega_{d-2}}{(d-1)G}\left(\frac{d-2}{4}\right)^{d-1}v^{d-1},\label{eq:vol}
\end{equation}
\begin{equation}
	\Phi=\left(\frac{\partial M}{\partial Q} \right)_{S,P}=\frac{4 \pi  Q r_+^{3-d}}{(d-3)\Omega_{d-2} G} = \frac{\pi  4^{d-2} Q (d-2)^{3-d}}{ (d-3)\Omega_{d - 2} G}v^{3-d},\label{eq:phi} 
\end{equation}
\begin{equation}
	T=\left( \frac{\partial M}{\partial S}\right)_{P,Q}=\frac{2^{-4 n} (d-2)^{2n-2} v^{2 n-1} }{8 \hat{\alpha}_n n }\left(128 P-\frac{\pi  16^d (d-2)^{4-2 d} Q^2 v^{4-2 d}}{\Omega_{d - 2}^2}\right)+\frac{8 (d-2) (d-2 n-1)}{8 (d-2)^2 n v \pi }\label{eq:temppp}
\end{equation}
 where $v={4r_+}/{\left(d-2\right)}$ denotes the specific volume. Also, using Eqs.\eqref{eq:S}, \eqref{eq:massp} and \eqref{eq:temppp}, entropy can be written as below 
\begin{equation}
	S= \frac{\hat{\alpha}_n(d-2)\Omega_{d - 2} n r_+^{d-2 n}}{4 G (d-2 n)}=\frac{ \hat{\alpha}_n (d-2)\Omega_{d - 2} n 4^{-d+2 n-1} ((d-2) v)^{d-2 n}}{G (d-2 n)},\label{eq:extS}
\end{equation}
 Hence, the equation of state can be found as
\begin{equation}
	P=P(T,v,Q,\hat{\alpha}_n)=\frac{\pi  2^{4 d-7}  Q^2 v^{4-2 d}}{(d-2)^{2 d-4}\Omega_{d - 2}^2}+\frac{ \hat{\alpha}_n v^{-2 n}}{16^{n-1}\pi(d-2)^{2 n-1}}\left( (d-2) n \pi T v-d+2 n+1 \right), \label{eq:eos}
\end{equation}
For four-dimensions $(d=4, n=1)$, Eq.\eqref{eq:eos} can be written as $P={2 Q^2}/{(\pi  v^4)}+{T}/{v}-{1}/{(2 \pi  v^2)}$. Now, we compare the pure Lovelock black holes thermodynamics behavior with van der Waals fluid.  For this aim, there must be an appropriate equation of state similar to that of van der Waals fluid  which is written as $P(\hat{T},\hat{v})$, where the superscript symbol, indicates the reduced parameter. The critical points $T_c,\ v_c \ \text{and} \ P_c$ where phase transition occurs can be obtained by using  the following relation
\begin{equation}
	 \partial_v P=\partial^2_{\ v}P=0, \label{eq:cric}
\end{equation}
  In Table \ref{tab:3} we have gathered the critical points  for different dimensions. The reduced parameters are defined as below
\begin{equation}
	\hat{T}=\frac{T}{T_c}, \ \ \ \ \ \hat{v}=\frac{v}{v_c}, \ \ \ \ \ \hat{P}=\frac{P}{P_c}. 
\end{equation}
 \begin{table}[t]
	\begin{center}
		\caption {\normalfont Critical points determined by Eq.\eqref{eq:cric}. }\label{tab:3}
		\begin{tabular}{|c|c|c|c|c|}
			\toprule
			 $n$	& $dim$   & $T_c$ & $ P_c $ & $v_c$   \\
			\midrule
             2&	6   &  $\frac{2 \sqrt{{2}/{3 \pi }} {\hat{\alpha}_2^{1/4}}}{15 {5^{1/4}} \sqrt{Q}}$  & ${\pi  \hat{\alpha}_2^2}/{270 Q^2}$ & $\frac{{5}^{1/4} \sqrt{{3Q}/{2 \pi }} }{{\hat{\alpha}_2^{1/4}}}$   \\
			\hline
             3	& 8  & $\frac{\sqrt{2} 3^{5/6} {\hat{\alpha}_3^{1/6}}}{35 {7^{1/6}} {(5 \pi Q)^{1/3}}}$ &${\pi ^3 \hat{\alpha}_3^2}/{1750 Q^2}$ &$\frac{\sqrt{2} {(5Q)^{1/3}} {7^{1/6}}}{3^{5/6} \pi ^{2/3} {\hat{\alpha}_3^{1/6}}}$  \\
			\hline
             4	&	10   &   $\frac{4\ 2^{7/8} {\hat{\alpha}_4^{1/8}}}{63 \sqrt{3} {(35 \pi Q )^{1/4}} }$  & ${32 \pi ^5 \hat{\alpha}_4^2}/{694575 Q^2}$ &$\frac{\sqrt{3} {(35Q)^{1/4}}}{2\ 2^{7/8} \pi ^{3/4} {\hat{\alpha}_4^{1/8}}}$   \\
			\bottomrule
		\end{tabular}
	\end{center}
\end{table}
In the following, we reach the reduced equation of state $\hat{P}=\hat{P}(\hat{T},\hat{v})$. We have obtained equation of state explicitly  for dimensions $d=6, \ d=8 \ \text{and} \ d=10$  as follows
\begin{align}
&\hat{P}_6=\frac{3}{5\hat{v}^8}-\frac{6}{\hat{v}^4}+\frac{32\hat{T}}{5\hat{v}^3}\label{eq:pv2},\\
&\hat{P}_8=\frac{5}{7\hat{v}^{12}}-\frac{10}{\hat{v}^6}+\frac{72\hat{T}}{7\hat{v}^5}\label{eq:pv3},\\
&\hat{P}_{10}=\frac{7}{9\hat{v}^{16}}-\frac{14}{\hat{v}^8}+\frac{128\hat{T}}{9\hat{v}^7}\label{eq:pv4}.
\end{align}
It is clear that the critical temperature and specific volume are $\hat{T}=1$ and $\hat{v}=1$, respectively, for all mentioned dimensions.  We have depicted the isotherm $\hat{P}-\hat{v}$ diagrams for various dimensions $d=6, \ d=8 \ \text{and} \ d=10$ in different temperatures $\hat{T}$, see Fig.\eqref{fig:p-v1}.  These   diagrams show that the thermodynamic behavior of pure Lovelock black holes as  thermodynamic systems are the same as the van der Waals fluid.  Furthermore, the temperature behavior $\hat{T}$ in various pressures $\hat{P}$, are depicted in Fig.\ref{fig:T-v1} with respect to the $\hat{v}$. It is seen that critical temperature $\hat{T}=1$, occurs at the critical specific volume $\hat{v}=1$ when the pressure gets the critical value $\hat{P}=1$. 

Now, let us investigate thermodynamic geometry of pure Lovelock black holes  in the extended phase space.
By taking Helmholtz free energy $F = U -T S$ as an appropriate thermodynamic potential, and $X^i=(T,V)$ as coordinates of thermodynamic manifold, the thermodynamic metric can be defined as follows
\begin{equation}
	dl^2=\frac{1}{T}\left(-\frac{\partial^2F}{\partial T^2}dT^2+\frac{\partial^2F}{\partial V^2}dV^2 \right) ,
\end{equation}
by using the differential form of free energy $dF=-SdT+PdV$  in the above relation we have
\begin{equation}
	dl^2=\frac{C_V}{T^2}dT^2+\frac{\left(\partial_V P \right)_T }{T}dV^2,\label{eq:dlex}
\end{equation}
where $C_V=T\left(\partial_TS \right)_V$ is the specific heat capacity at the constant volume $V$ and  as  before, thermodynamic Ricci scalar diverges at phase transition points. Using metric in Eq.\eqref{eq:dlex} the Ricci scalar can be easily found as 
\begin{equation}
	R=\frac{\left(\partial_V P \right)^2-T^2\left(\partial_{T,V} P \right)^2+2T^2\left(\partial_V P \right)\left(\partial_{T,T,V} P \right)}{2C_V\left(\partial_V P \right)^2},
\end{equation}
Considering Eqs.\eqref{eq:pv2}-\eqref{eq:pv4}, it was found that the equation of state depends  on temperature $T$ linearly and as a result we have $\left(\partial_{T,T,V} P \right)=0$, therefore the above scalar curvature relation reduces to the following form
\begin{equation}
	R=\frac{1}{2C_V}\left( 1-\left(T\frac{\partial_{V,T} P }{\partial_V P} \right)^2 \right) .\label{eq:Rcv}
\end{equation}
We can also obtain extrinsic curvature in thermodynamic geometry. For the metric in Eq.\eqref{eq:dlex}  extrinsic curvature can be written as follows \cite{mans_crit}
\begin{equation}
	K=\frac{1}{2\sqrt{C_V}}\left( 1-T\frac{\partial_{V,T}P}{\partial_VP}\right) .\label{eq:kcv}
\end{equation}
 The specific heat capacity of van der Waals fluid at constant volume is $C_V=3/2 \ k_B$. Note that Eq.\eqref{eq:extS} indicates  that $C_V=T(\partial S/\partial T)_{r_+}=0$ and therefore in the following we use normalize form of curvatures as $R_N=R C_V$ and $K_N=K\sqrt{C_V}$. In the following, by using critical points reported in Table \ref{tab:3} and the fact that $V_c=\Omega_{d - 2}/(d-1)\left((d-2)/4 \right)^{d-1}v_c^{d-1}$, we rewrite the equations of state in Eqs.\eqref{eq:pv2},\eqref{eq:pv3} and \eqref{eq:pv4} in the  form of $\hat{P}=\hat{P}(\hat{T},\hat{V})$ as below
\begin{align}
	&\hat{P}_6=\frac{3}{5\hat{V}^{8/5}}-\frac{6}{\hat{V}^{4/5}}+\frac{32\hat{T}}{5\hat{V}^{3/5}},\\
	&\hat{P}_8=\frac{5}{7\hat{V}^{12/7}}-\frac{10}{\hat{V}^{6/7}}+\frac{72\hat{T}}{7\hat{V}^{5/7}},\\
	&\hat{P}_{10}=\frac{7}{9\hat{V}^{16/9}}-\frac{14}{\hat{V}^{8/9}}+\frac{128\hat{T}}{9\hat{V}^{7/9}}.
\end{align}
By   using above equations of state in Eqs.\eqref{eq:Rcv} and \eqref{eq:kcv}, one can evaluate normalized
 Ricci scalar and extrinsic curvature relations for $d=6, \ d=8 \ \text{and} \ d=10$ as follows
\begin{align}
& {6d:}  \ \ \ R_{N}=-\frac{\left(5 \hat{V}^{4/5}-1\right) \left(8 \hat{T} \hat{V}-5 \hat{V}^{4/5}+1\right)}{2 \left(4 \hat{T} \hat{V}-5 \hat{V}^{4/5}+1\right)^2}, \ \ \ \ \ \ \ \ \ \ K_{N}=\frac{1-5 \hat{V}^{4/5}}{8 \hat{T} \hat{V}-10 \hat{V}^{4/5}+2},\\
& {8d:} \ \ \ R_{N}=-\frac{\left(7 \hat{V}^{6/7}-1\right) \left(12 \hat{T} \hat{V}-7 \hat{V}^{6/7}+1\right)}{2 \left(6\hat{T} \hat{V}-7 \hat{V}^{6/7}+1\right)^2}, \ \ \ \ \ \ \ \ K_{N}=\frac{1-7 \hat{V}^{6/7}}{12 \hat{T} \hat{V}-14 \hat{V}^{6/7}+2},\\
& {10d:} \ \ \ R_{N}=-\frac{\left(9 \hat{V}^{8/9}-1\right) \left(16 \hat{T} \hat{V}-9 \hat{V}^{8/9}+1\right)}{2 \left(8 \hat{T} \hat{V}-9 \hat{V}^{8/9}+1\right)^2}, \ \ \ \ \ \ K_{N}=\frac{1-9 \hat{V}^{8/9}}{16 \hat{T} \hat{V}-18 \hat{V}^{8/9}+2}.
\end{align}
\begin{figure}[t]
	\centering
	\begin{subfigure}{.3\textwidth}
		\centering
		\includegraphics[width=0.9\textwidth]{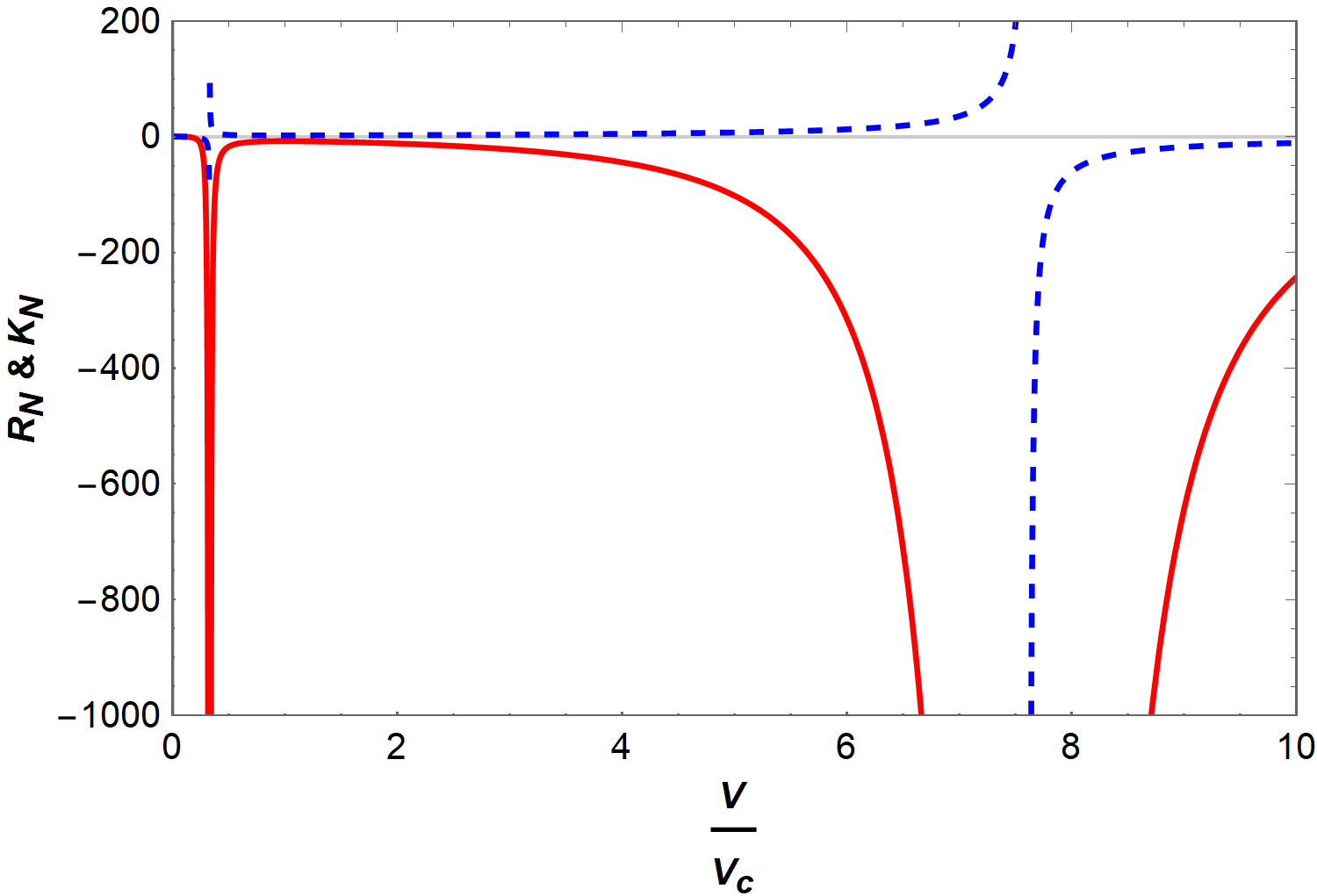}
		\caption{}
	\end{subfigure}%
	\begin{subfigure}{.3\textwidth}
		\centering
		\includegraphics[width=0.9\textwidth]{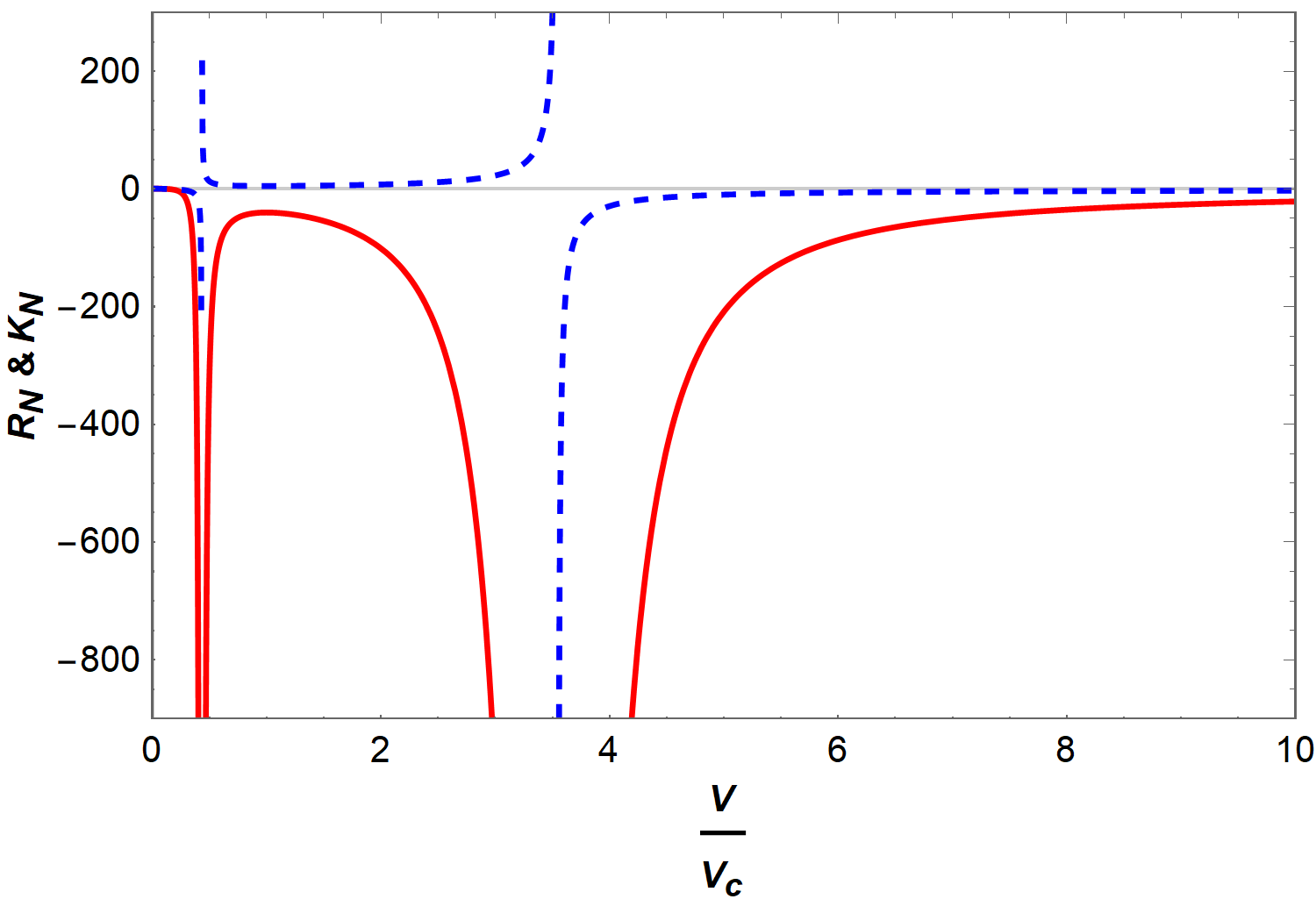}
		\caption{}
	\end{subfigure}%
	\begin{subfigure}{.3\textwidth}
		\centering
		\includegraphics[width=0.9\textwidth]{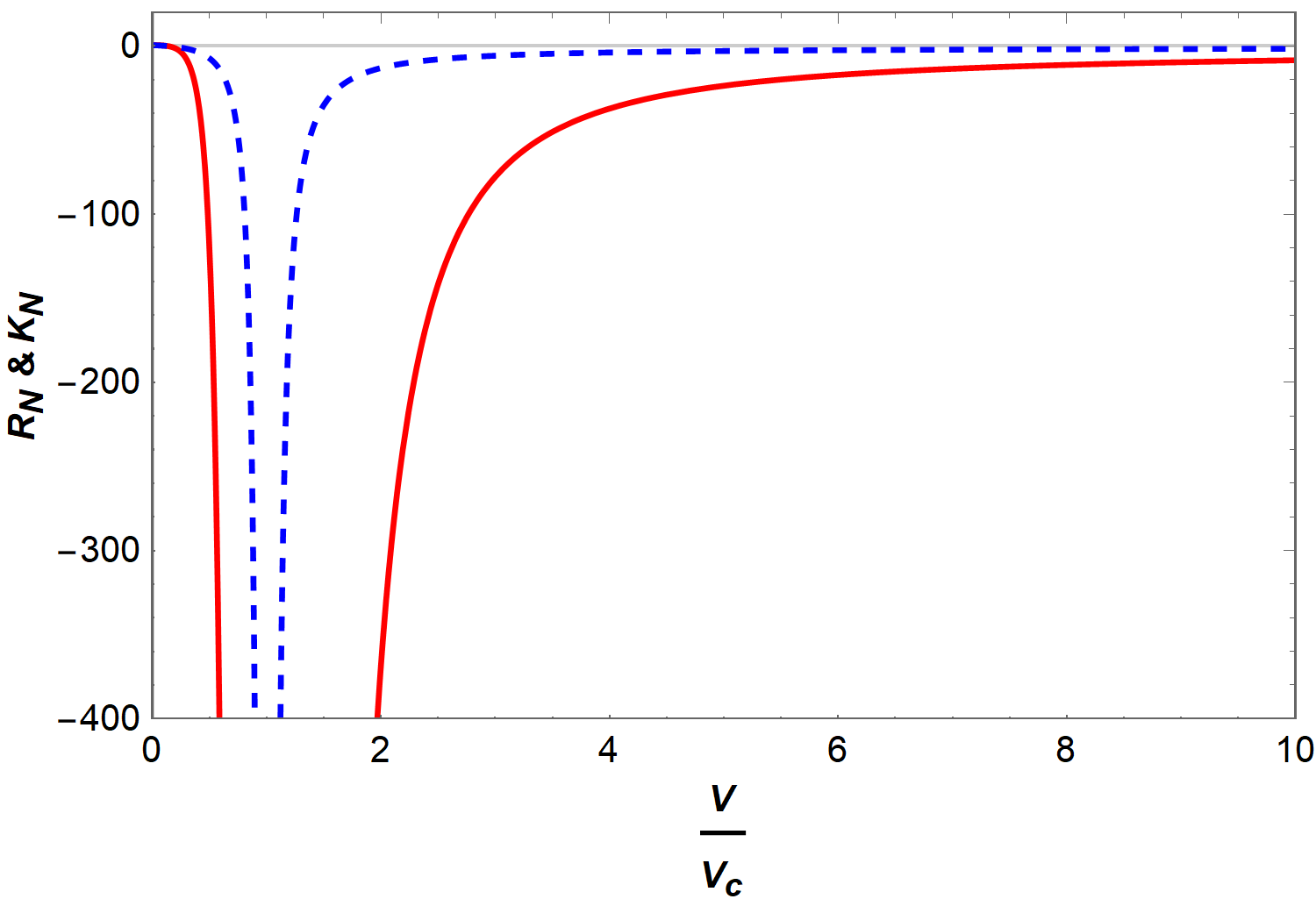}
		\caption{}
	\end{subfigure}
	\begin{subfigure}{.3\textwidth}
		\centering
		\includegraphics[width=0.9\textwidth]{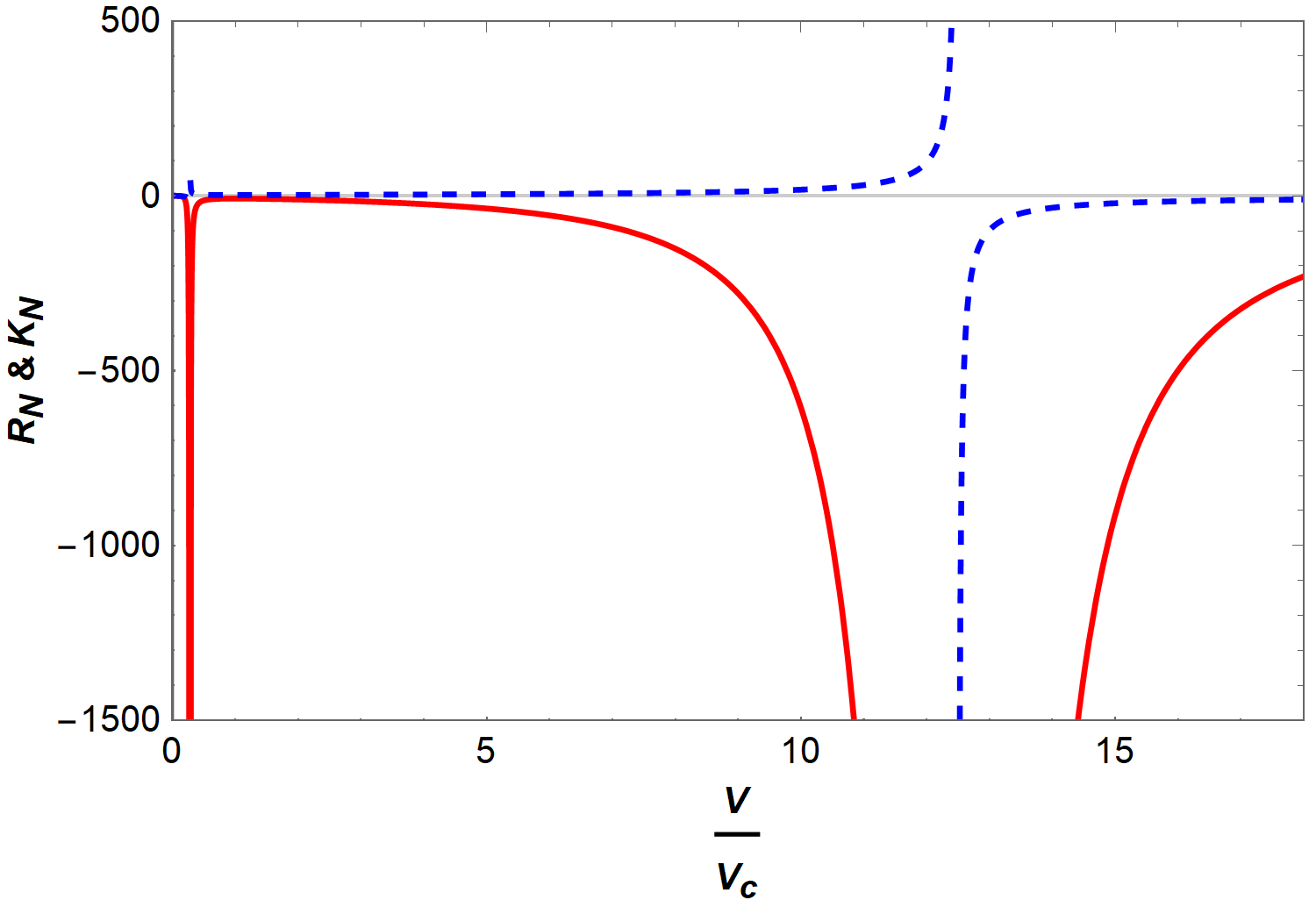}
		\caption{}
	\end{subfigure}%
	\begin{subfigure}{.3\textwidth}
		\centering
		\includegraphics[width=0.9\textwidth]{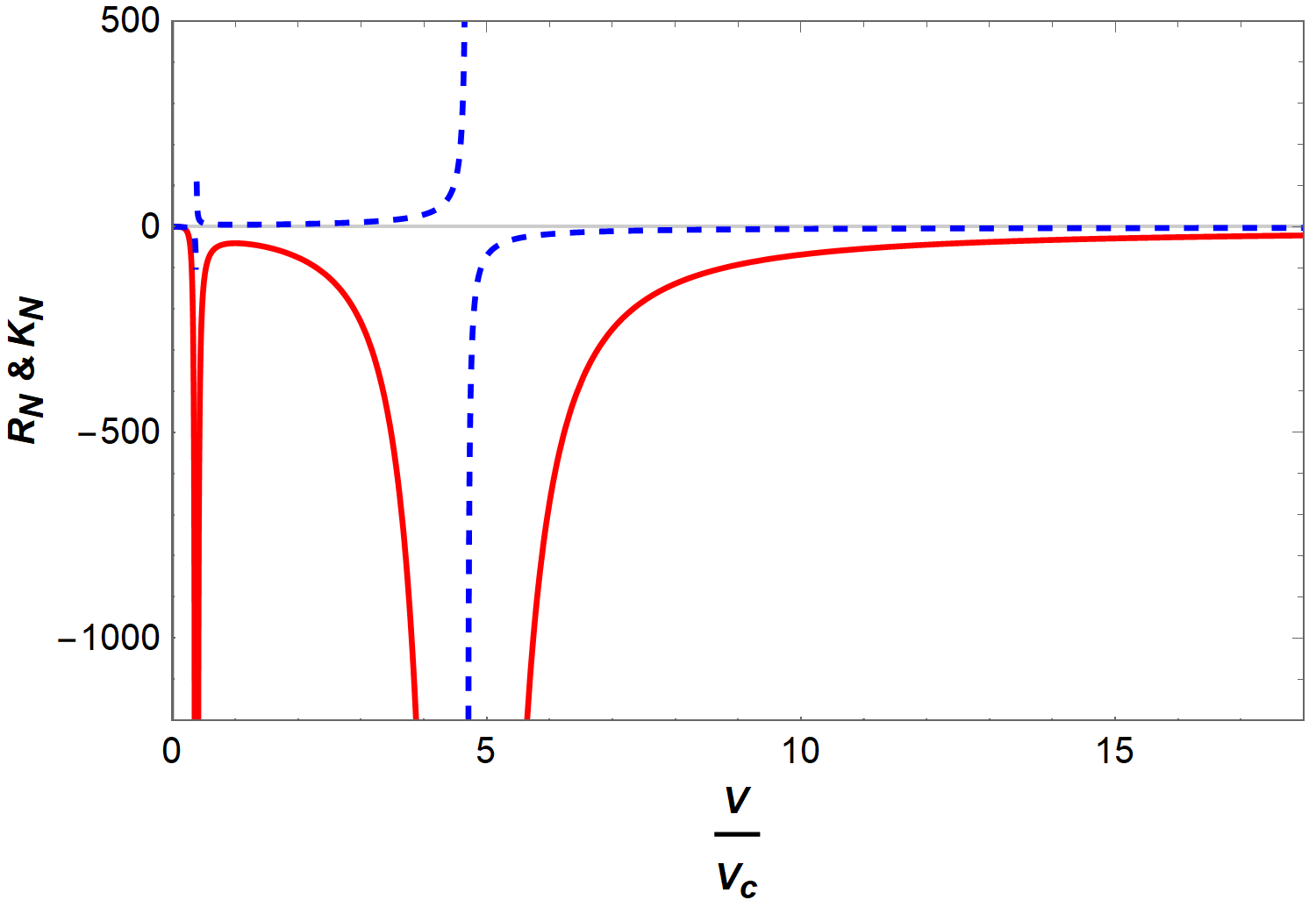}
		\caption{}
	\end{subfigure}%
	\begin{subfigure}{.3\textwidth}
		\centering
		\includegraphics[width=0.9\textwidth]{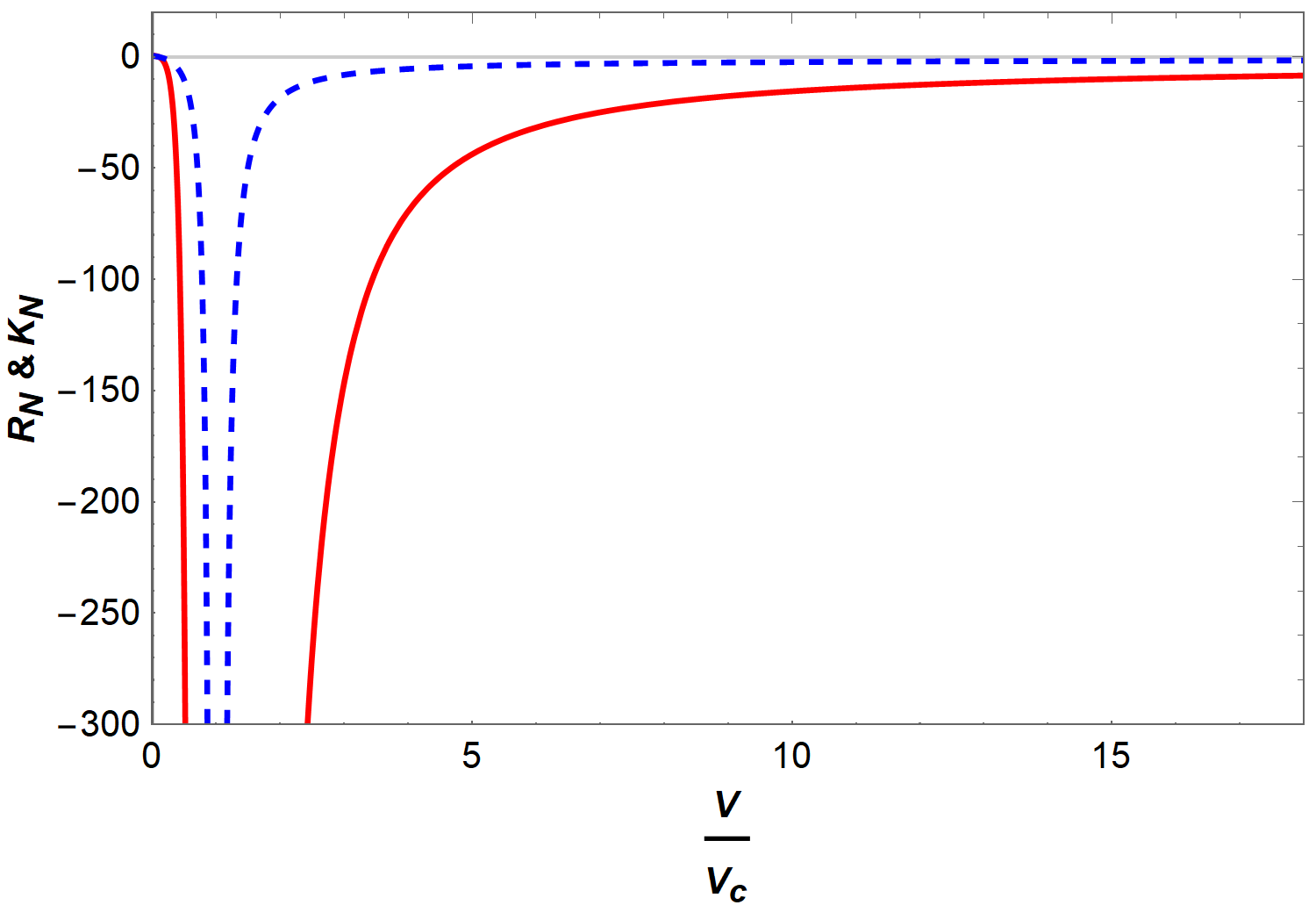}
		\caption{}
	\end{subfigure}
	\begin{subfigure}{.3\textwidth}
		\centering
		\includegraphics[width=0.9\textwidth]{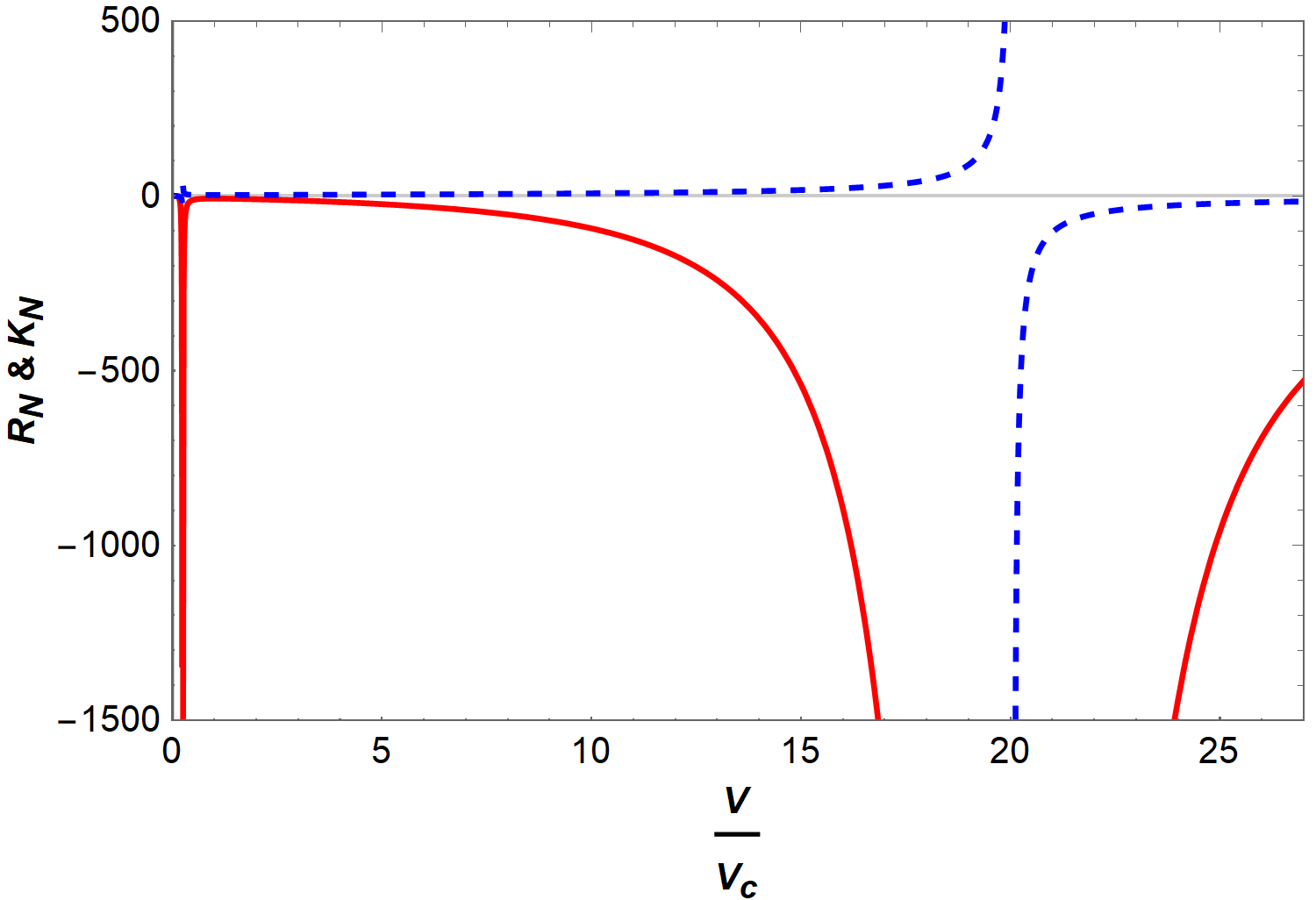}
		\caption{}
	\end{subfigure}%
	\begin{subfigure}{.3\textwidth}
		\centering
		\includegraphics[width=0.9\textwidth]{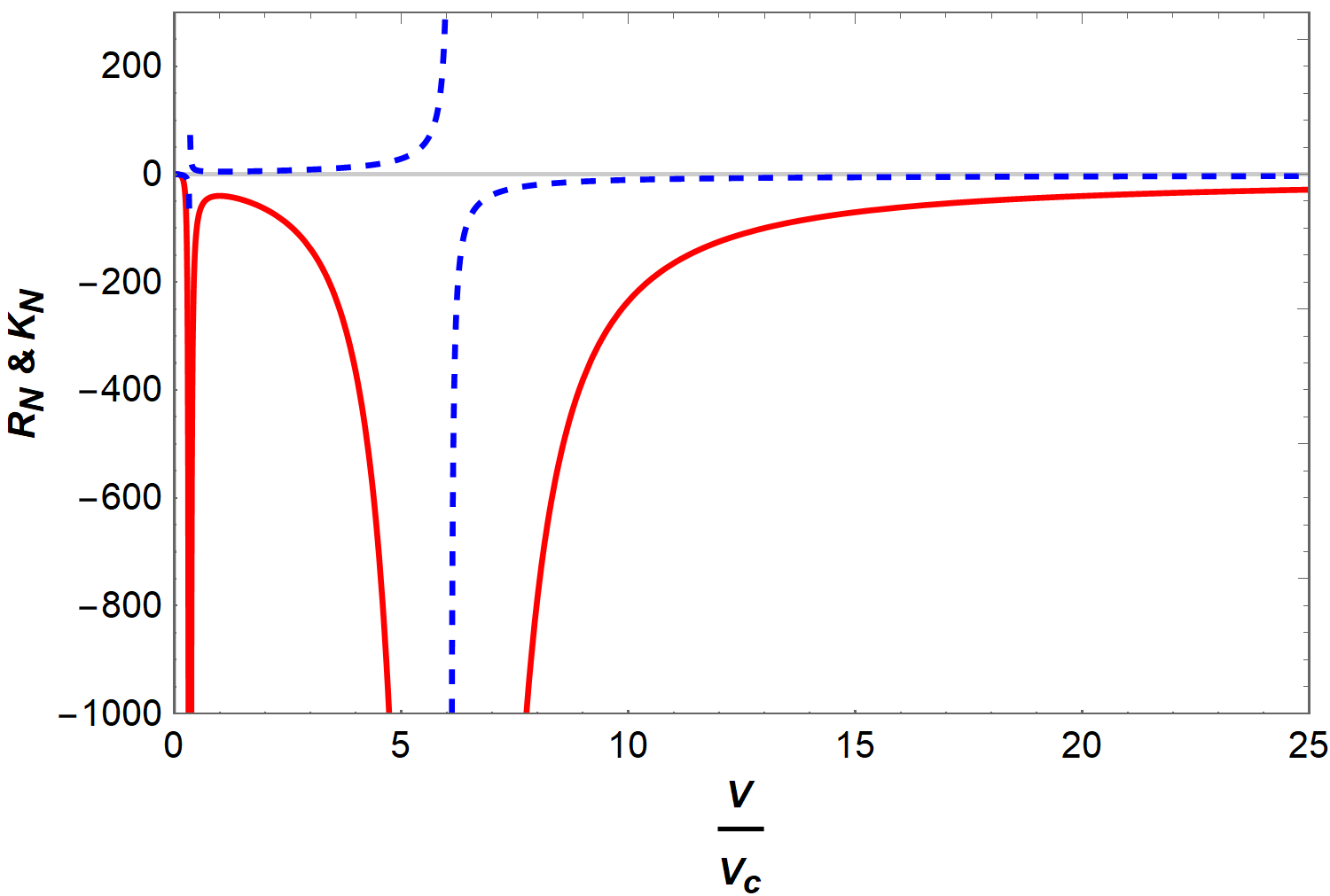}
		\caption{}
	\end{subfigure}
	\begin{subfigure}{.3\textwidth}
		\centering
		\includegraphics[width=0.9\textwidth]{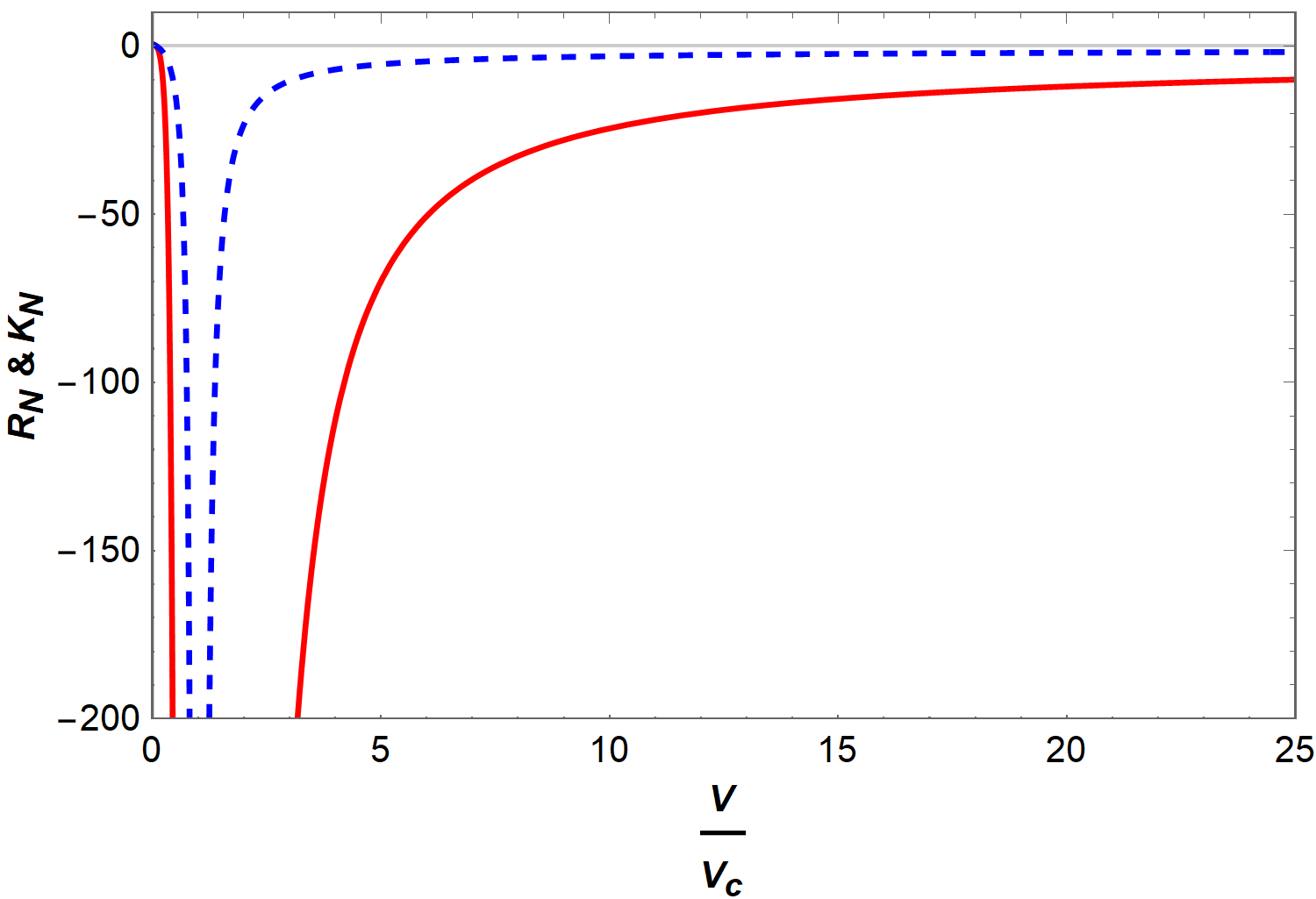}
		\caption{}
	\end{subfigure}%
	\caption{\normalfont Graphs of the normalized thermodynamic Ricci scalar (solid red curve) and extrinsic curvature (dashed blue curve) as a function of thermodynamic volume $\hat{V}$.   {\itshape Top}: {\itshape$6 \ dimensions$}   for $ (a) \ \hat{T}=0.8, \  (b)\ \hat{T}=0.9, \  (c) \   \hat{T}=1$.  {\itshape Middle}: {\itshape$8 \ dimensions$} for $ (d) \ \hat{T}=0.8, \  (e) \ \hat{T}=0.9, \  (f) \ \hat{T}=1$.  {\itshape Bottom}: {\itshape$10 \ dimensions$} for $ (g) \ \hat{T}=0.8, \  (h) \ \hat{T}=0.9, \  (i) \  \hat{T}=1$.}
	\label{fig:extRicci}
\end{figure}
We have plotted normalized thermodynamic Ricci scalar and the extrinsic curvature with respect to the thermodynamic volume $\hat{V}$ for various dimensions in Fig \ref{fig:extRicci}. As in previous sections, the  phase transition occurs exactly at singularities of scalar curvature and we expect a repulsive interaction between microstructure of pure Lovelock black holes associated with $R>0$. It is seen  in Fig \ref{fig:extRicci}, that there are two critical points where  $R_N$ and $K_N$ diverge,  the first point   exists at  small $\hat{V}$ and the second one is at large $\hat{V}$. The two divergent points appear at fixed low temperatures as $\hat{T}=0.75, \ \hat{T}=0.95$  for $ d=6, \ d=8 \ \text{and} \ d=10$ in $a, \ b, \ d, \ e, \ g, \ h$ in Fig.\ref{fig:extRicci}. Furthermore, as $T$ tends to the critical value $(T=1)$, the two critical points tends to each other and finally coincide as in Fig. \ref{fig:extRicci}. Note that for higher values of temperature, $ \hat{T}>1 $, there is no critical point. 

Now, we examine the critical behavior of these black holes using the Gibbs free energy ($G=M-TS$). Using Eqs. \eqref{eq:massp}, \eqref{eq:temppp} and \eqref{eq:extS}, Gibbs free energy may be written as below
\small
\begin{equation}
	G=\frac{\Omega_{d - 2} 4^{-d-1} (d-2)^{d-1} v^{d-1} \left(\hat{\alpha}_n (d-1) 16^n (d-2)^{1-2 n} v^{-2 n}+16 \pi  (1-2 n) P\right)}{\pi  (d-1) (d-2 n)}+\frac{\pi  2^{2 d-5} (d-2)^{3-d} Q^2 (2 d-2 n-3) v^{3-d}}{\Omega_{d - 2} (d-3) (d-2 n)}.
\end{equation}
\normalsize
In the following, by  using  reduced parameters we  omit the electric charge $Q$, and the coupling constant $\hat{\alpha}_n$. Therefore,    it  becomes  similar to van der Waals fluid in term of $\hat{v}$ and $\hat{P}$ as below
\begin{align}
&\hat{G}_6=-\frac{ 3 \hat{P}}{32} \hat{v}^5+\frac{5}{32 \hat{v}^3}+\frac{15 \hat{v}}{16},\\
&\hat{G}_{8}=-\frac{5 \hat{P}}{72}  \hat{v}^7+\frac{7}{72 \hat{v}^5}+\frac{35 \hat{v}}{36},\\
&\hat{G}_{10}=-\frac{7 \hat{P}}{128}  \hat{v}^9+\frac{9}{128 \hat{v}^7}+\frac{63 \hat{v}}{64}.
\end{align} 
\begin{figure}[t]
	\centering
	\begin{subfigure}{.35\textwidth}
		\centering
		\includegraphics[width=1\textwidth]{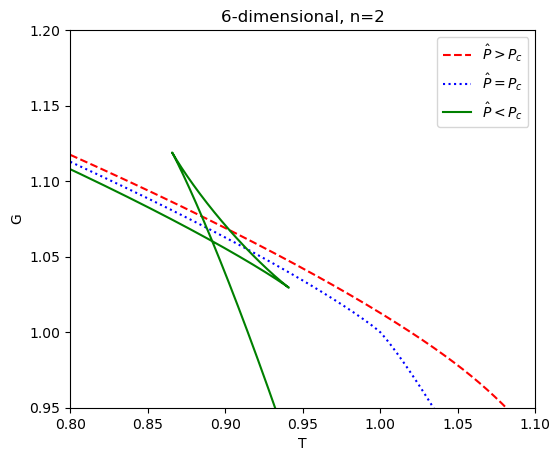}
		\caption{}
	\end{subfigure}%
	\begin{subfigure}{.35\textwidth}
		\centering
		\includegraphics[width=1\textwidth]{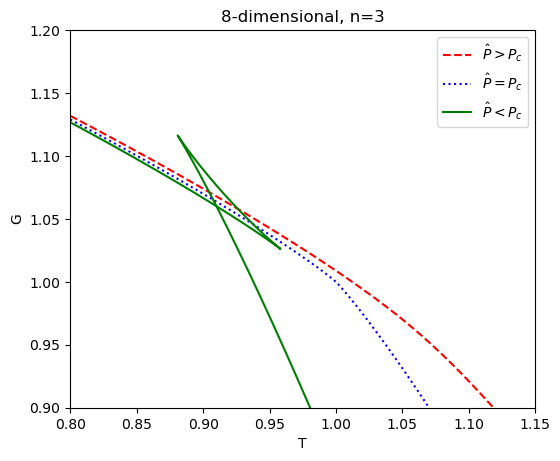}
		\caption{}
	\end{subfigure}
	\begin{subfigure}{.35\textwidth}
		\centering
		\includegraphics[width=1\textwidth]{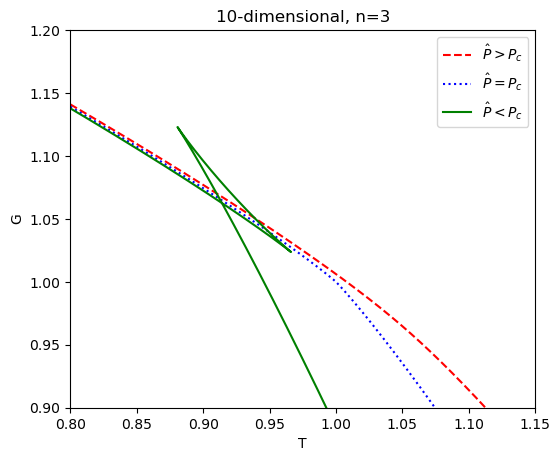}
		\caption{}
	\end{subfigure}
	\caption{\normalfont Gibbs free energy as a function of temperature $\hat{T}$, in various dimensions $d=6, \ d=8 \ \text{and} \ d=10$ for different pressures which are larger (dashed red curves), smaller (green curves) than the critical pressure and at the critical pressure (blue dot curves). (a)  $6 \ dimensions$  for  $\hat{P}=1.5 \ \hat{P}=1 , \ \text{and} \ \hat{P}=0.5  $. (b) $8 \ dimensions$  for $\hat{P}=1.8, \ \hat{P}=1, \ \text{and} \ \hat{P}=0.4 $. (c)   $10 \ dimensions$  for  $\hat{P}=2, \ \hat{P}=1, \ \text{and} \ \hat{P}=0.3 $. }
	\label{fig:gibbs}
\end{figure}
The Gibbs free energies $\hat{G}$ as functions of $\hat{T}$  are depicted in Fig.\ref{fig:gibbs}. As we expected, for $\hat{P}<1$ the diagrams show swallowtail behavior which indicates a phase transition from small  to large black holes. For interpretation of the cosmological constant as a thermodynamic variable in AdS/CFT see \cite{cft1,cft2, cft3, cft4, cft5, cft6, cft7, cft8, cft9, cft10,cft11} 
\section{CONCLUSIONS}\label{sec:conc}
We explored the thermodynamic characteristics of black holes in pure Lovelock gravity in this work. We studied thermodynamic geometry and derived the thermodynamic scalar and extrinsic curvatures.  We compared the specific heat capacity with the thermodynamic curvature and depicted the associated diagrams.  Our results show  the exact correspondence between thermodynamic Ricci scalar $R $ and specific heat $C_Q$ at critical point.  According to the thermodynamic Ricci scalar  for the following horizon radius, $3/2<r_+<5/2$, $3/2\le r_+<2$ and $1.4<r_+<3/2$ in $d=6$, $d=8$ and $d=10$, respectively,  we have attractive interactions ($R<0$) for microstates which means a bosonic behavior in this range of parameters. There is repulsive interactions ($R>0$) or in other words fermionic behavior for $r_+>2$, $r_+>1.5$, $r_+>1.2$, $r_+\geq1.2$ in $d=4$, $d=6$, $d=8$ and $d=10$, respectively. It was seen that extrinsic curvature has the  same behavior as the specific heat capacity. Moreover,  we investigated the critical behavior of thermodynamic curvatures near the critical points  to reach the critical exponents and amplitudes. We found the critical exponent $\alpha=0$ in this model which is similar to the  van der Waals fluid.  We analyzed phase transition by Ehrenfest approach and concluded the validity of the Ehrenfest's equations in pure Lovelock black holes that we have second order phase transition. Finally, we investigated pure Lovelock black holes in the extended phase space. We derived equations of state for various dimensions and investigated the critical behavior, too. At the critical temperature $\hat{T}=\hat{T}_c$ a phase transition occurs from small to a large black holes which is  similar to phase  transition between gas and liquid in the van der Waals fluid. Furthermore, we depicted the isotherm and isobaric diagrams for pure Lovelock black holes for $d=6$, $d=8$ and $d=10$ and found a similar behavior to the van der Waals fluid. Furthermore, we showed that there are two critical points where the normalized Ricci scalar $R_N$ and normalized extrinsic curvature $K_N$ diverge. However at the critical temperature we only observe one singular point. Also, we depicted  Gibbs free energy for our thermodynamic system which  represents swallowtail behavior. The phase transition from small to large black holes is a first order one. 
\section*{Acknowledgments}
We would like to thank Naresh Dadhich, and Seyed Ali Hosseini Mansoori for comments and discussions. 
\appendix
\section*{Appendix A}\label{App:A}
In the following we review some useful identities. Consider the functions $f(x,y),g(x,y)$ and $h(x,y)$, we have the following identity
\begin{equation}
\left(\frac{\partial f}{\partial g}\right)_h=\frac{\{f,h\}_{x,y}}{\{g,h\}_{x,y}},\tag{A.1} \label{eq:A.1}
\end{equation}
where $\{f,h\}_{x,y}$ is Numba bracket that is defined by the following equation
\begin{equation}
\{f,h\}_{x,y}=\left|\begin{matrix}\left(\frac{\partial f}{\partial x}\right)_y&\left(\frac{\partial f}{\partial y}\right)_x\\\left(\frac{\partial h}{\partial x}\right)_y&\left(\frac{\partial h}{\partial y}\right)_x\\\end{matrix}\right|=\left(\frac{\partial f}{\partial x}\right)_y\left(\frac{\partial h}{\partial y}\right)_x-\left(\frac{\partial f}{\partial y}\right)_x\left(\frac{\partial h}{\partial x}\right)_y.\tag{A.2} \label{eq:A.2}
\end{equation}
In another case, assume that $f$, $g$, $h_1$ and $h_2$ are functions of $x,y,z$,the following identity was introduced in \cite{hessian}
\begin{equation}
\left(\frac{\partial f}{\partial g}\right)_{h_1,h_2}=\frac{\{f,h_1,h_2\}_{x,y,z}}{\{g,h_1,h_2\}_{x,y,z}},\tag{A.3}
\end{equation}
where the generalized Numba bracket is defined as follows
\begin{equation}
\{f,h_1,h_2\}_{x,y,z}=\left|\begin{matrix}\left(\frac{\partial f}{\partial x}\right)_{y,z}&\left(\frac{\partial f}{\partial y}\right)_{x,z}&\left(\frac{\partial f}{\partial z}\right)_{x,y}\\\left(\frac{\partial h_1}{\partial x}\right)_{y,z}&\left(\frac{\partial h_1}{\partial y}\right)_{x,z}&\left(\frac{\partial h_1}{\partial z}\right)_{x,y}\\\left(\frac{\partial h_2}{\partial x}\right)_{y,z}&\left(\frac{\partial h_2}{\partial y}\right)_{x,z}&\left(\frac{\partial h_2}{\partial z}\right)_{x,y}\\\end{matrix}\right|.
\label{eq:A.4} \tag{A.4}
\end{equation}
To generalize \eqref{eq:A.4}, assume the functions $f(q_1,q_2,...,q_{m+1})$ , $g(q_1,q_2,...,q_{m+1})$ and $h_{1,2,...,m}(q_1,q_2,...,q_{,m+1})$, the the following identity was proved in \cite{hessian}
\begin{equation}
\left(\frac{\partial f}{\partial g}\right)_{h_1,h_2,...,h_m}=\frac{\{f,h_1,h_2,...,h_m\}_{q_1,q_2,...,q_{m+1}}}{\{g,h_1,h_2,...,h_m\}_{q_1,q_2,...,q_{m+1}}},\tag{A.5}\label{eq:A.5}
\end{equation} 
where $\{f,h_1,h_2,...,h_m\}$ is defined as follows
\begin{equation}
\{f,h_1,h_2,...,h_m\}_{q_1,q_2,...,q_{m+1}}=\sum_{ijk...n=1}^{m+1}\epsilon_{ijk...n}\frac{\partial f}{\partial q_i}\frac{\partial h_1}{\partial q_j}\frac{\partial h_2}{\partial q_k}...\frac{\partial h_m}{\partial q_n}.\tag{A.6}\label{eq:A.6}
\end{equation}

\end{document}